\documentclass[a4paper,english,twocolumn,preprintnumbers,amsmath,amssymb,superscriptaddress]{revtex4}
\pdfoutput=1
\usepackage[T1]{fontenc}
\usepackage[latin1]{inputenc}
\usepackage{babel}
\usepackage{array}
\usepackage{graphicx}

\makeatletter
\newcommand{\ppi}{p_i}

\begin{document}

\title{Automatic, optimized interface placement in forward flux sampling simulations}

\author{Kai Kratzer}
\affiliation{ICP, Institute for Computational Physics, University of Stuttgart, 
Allmandring 3, 70569 Stuttgart, DE.}

\author{Axel Arnold}
\affiliation{ICP, Institute for Computational Physics, University of Stuttgart, 
Allmandring 3, 70569 Stuttgart, DE.}

\author{Rosalind J. Allen}
\affiliation{SUPA, School of Physics, The University of Edinburgh, 
Mayfield Road, Edinburgh EH9 3JZ, UK.}

\date{\today}

\begin{abstract}
Forward flux sampling (FFS) provides a convenient and efficient way to simulate rare events in equilibrium or non-equilibrium systems. FFS ratchets the system from an initial state to a final state via a series of interfaces in phase space. The efficiency of FFS depends sensitively on the positions of the interfaces. 
We present two alternative methods for placing interfaces automatically and adaptively in their optimal locations, on-the-fly as an FFS simulation progresses, without prior knowledge or user intervention. 
These methods allow the FFS simulation to advance efficiently through bottlenecks in phase space by placing more interfaces where the probability of advancement is lower. The methods are demonstrated  both for a single-particle test problem and for the  crystallization of Yukawa particles. 
By removing the need for manual interface placement, our methods both facilitate the setting up of FFS simulations and  improve their performance,  especially for rare events which involve complex trajectories through phase space, with many bottlenecks. 

\end{abstract}

\maketitle

\newpage

\section{Introduction}\label{sec:intro}
Many important processes in nature can be described as  rare
events -- i.e. events that happen rapidly but unpredictably, with long waiting times between occurrences. Examples of such processes range from large-scale problems such as electricity or computer network failures,  to molecular level processes such as the formation of crystal nuclei or vapour bubbles in metastable liquids. Rare events are difficult to study, either in experiments or in computer simulations, because most of the observation time is spent waiting for the fluctuation-driven event to happen. In simulations, this problem can be overcome using rare event simulation techniques such as umbrella sampling~\cite{umbrella1,daan}, Bennett-Chandler methods~\cite{bencha1,bencha2,daan}, transition path sampling~\cite{dellago1,dellago2,bolhuis_arpc}, transition interface  sampling~\cite{vanerp1,vanerp2,vanerp2008}, milestoning~\cite{faradjian04, west07, ve08}, nudged elastic 
band~\cite{henkelman1,henkelman2}, string methods~\cite{Ren02,e_05}, weighted-ensemble methods~\cite{huber1996}, non-equilibrium umbrella sampling or forward flux sampling (or splitting)-type methods~\cite{fFFS,FFS2,efficiency,ffs,barrier,borrero2009,borrero2007,dinner1,dinner2,escobedorev}.
All of these methods aim to enhance the sampling in the region of phase space (or trajectory space) that corresponds to the rare event, while reducing the amount of time the simulation spends in the uninteresting phase space (or trajectory space) regions corresponding to the waiting times. 

In this paper, we focus on the forward flux sampling (FFS) approach \cite{ffs}. In FFS, one uses an order parameter to measure the progress of the system from the initial state towards the final state.
The region of phase space between the initial and final states is partitioned by a series of  interfaces defined by specific values of the order parameter. These interfaces are used to ratchet the system from the initial to the final state. Short trajectories are fired from the initial state; if these reach the first interface, they are used as starting points for further trajectories, which, if they reach the second interface, are used as starting points for further trajectories, etc.  During this procedure, the fraction of trajectories which reach the next interface is monitored. The product of these ``success probabilities'' over all interfaces, together with the flux of trajectories out of the initial state, gives the transition rate from initial to final state. Unbiased transition trajectories  can be reconstructed from the collection of trajectories between interfaces. FFS provides a convenient way to simulate rare events in stochastic dynamical systems, because it is rather simple to implement and 
allows direct calculation of the transition rate. 
Importantly, FFS  is suitable for both equilibrium and non-equilibrium systems (since it does not require {\em{a priori}} knowledge of the phase space density) \cite{ffs}. Recent advances in FFS-type methods include the development of different algorithms for the trajectory-firing procedure~\cite{FFS2}, computation of phase-space densities as well as transition rates~\cite{valeriani}, analysis and optimization of the efficiency of the method \cite{efficiency,ffs,borrero2007,borrero2009,escobedorev}, and the development of FFS-like methods for systems which are out of the stationary state \cite{spres,nsffs}. While FFS is of course not a panacea for all rare event problems \cite{sear,vanerp}, it is widely and successfully used for a range of systems, some of which would be difficult or impossible to tackle with other methods. The validity of FFS has been extensively tested against brute force simulations and other rare event simulation methods for a range of problems \cite{fFFS,ffs,dijkstra,valeriani_nuc,vega,
borrero_lattice}

In this paper, we focus on the placement of interfaces in FFS. The  number of interfaces and their positions are important inputs in FFS, since poor interface placement can have strongly detrimental effects on the efficiency \cite{efficiency,borrero}. If the interfaces are placed too far apart the probability of reaching the next interface will be very low, and much effort will be wasted firing trajectories that fail to progress. On the other hand,  if  the interfaces are too close together, trajectories will be highly correlated between successive interfaces so that little new information is gained at each interface.

Borrero and Escobedo \cite{borrero,borrero2} have shown that, for a fixed number of interfaces, the efficiency of FFS simulations is optimized when the flux of trajectories between interfaces is equalized: i.e., for a fixed number of trial trajectories per interface, the probability of reaching the next interface should be equal for all interfaces. 
This criterion allows optimal interface placement, if one has prior knowledge of how the success probabilities depend on the order parameter.
Such knowledge is, however, not usually available. Borrero and Escobedo suggest beginning with non-optimized interfaces and using the constant flux criterion to iteratively improve the interface placement in successive FFS runs \cite{borrero,borrero2}. While this is a good strategy for some problems, it is problematic for computationally expensive systems with high barriers. For these systems, FFS simulations with poorly chosen interface sets simply will not finish in a reasonable computational time. This forces the user to spend much effort on finding a reasonable initial interface set, by manual trial-and-error. Moreover, repeating the FFS simulations to obtain iteratively better interface sets is  computationally expensive. To our knowledge, the only interface-based method that does not require {\em{a priori}} interface placement is adaptive multi-level splitting (AMS) \cite{ams}. In this method, successive interfaces are placed adaptively, based on the furthest point in order parameter space reached by 
previous trajectories. Practical implementation of AMS is, however, more complex than for standard FFS, because one needs to keep track of the histories of previous trajectories in order to determine the start points of new trajectories. This is likely to involve coding and storage overheads, particularly for large systems.

In this paper, we present two methods which allow optimal placement of interfaces in standard FFS simulations, on-the-fly, {\em{without  any prior knowledge}}. We first use theoretical arguments to estimate the optimal range for the flux between interfaces, when the number of interfaces is not fixed (section \ref{sec:bg}). In section \ref{sec:ipm} we  present our algorithms and discuss the situations in which we expect each to be advantageous.  
We demonstrate the performance of both methods in section \ref{sec:simex}, first   for a one-particle test problem and then for a computationally expensive rare event problem: crystallization in a system of Yukawa particles. Finally we present our conclusions in section \ref{sec:dis}.

\section{Optimization principles for interface placement in FFS}\label{sec:bg}

In this section, we first briefly describe the ``direct'' FFS algorithm. We then review the work of Borrero and Escobedo 
which shows that, for a fixed number of interfaces, optimal efficiency requires  equal fluxes between interfaces \cite{borrero,borrero2}. Building on this work, we  
establish the optimal range for the transition probability between interfaces, in the case where the number of interfaces is not constrained. This optimal range will be used as input for the computational algorithms described in section \ref{sec:ipm}.

\subsection{The direct FFS algorithm (DFFS)}\label{sec:ffs}
\begin{figure}
\begin{center}
{\rotatebox{0}{{\includegraphics[scale=0.33,clip=true]{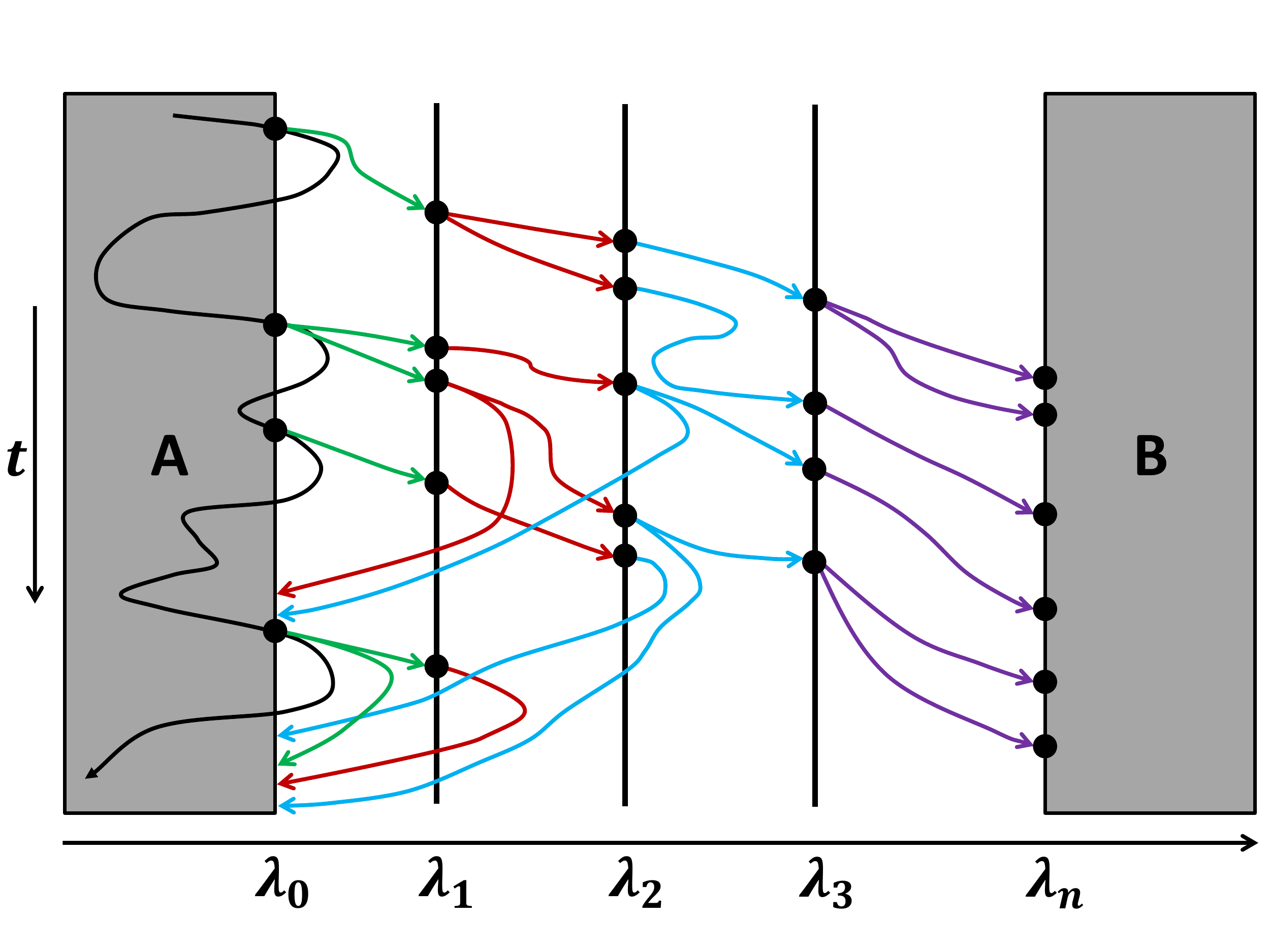}}}}
\caption{Schematic illustration of the DFFS algorithm. The barrier region between $\lambda_A$ and $\lambda_B$ is partitioned by a series of interfaces, defined as values of the order parameter $\lambda$ (horizontal axis). The black dots denote stored configurations; and the coloured arrows (colour coded by interface) represent trajectories. The vertical axis denotes simulation time, increasing downwards.
  \label{fig:ffs_fig}}
\end{center}
\end{figure}

The aim of FFS is to compute the transition rate $k_{AB}$ from an initial state $A$ to a final state $B$, while at the same time sampling the associated transition trajectories. The transition rate $k_{AB}$ is given by $k_{AB} = \Phi P_B $ \cite{vanerp1}, where $\Phi$ is the flux of trajectories leaving the initial state, and  $P_B$ is the probability that a trajectory that leaves the initial state will subsequently make it to the final state (rather than returning to the initial state). In FFS, the initial and final states are defined in terms of an order parameter $\lambda$, such that if $\lambda < \lambda_A$ the system is in the initial state and if $\lambda > \lambda_B$ it is in the final state. Intermediate values of $\lambda$ ($\lambda_A < \lambda < \lambda_B$) correspond to the ``barrier'' region. This barrier region is partitioned by a series of $n$ interfaces, defined by specific values of $\lambda$, such that $\lambda_i < \lambda_{i+1}$,  $\lambda_0 \equiv \lambda_A$ and $\lambda_n \equiv \lambda_
B$ (
see Figure \ref{fig:ffs_fig}).
The probability $P_B$ can be written as \cite{vanerp1}
\begin{equation}\label{eq:ppi}
P_B = \prod_{i=0}^{n-1}\ppi
\end{equation}
where $\ppi$ is the conditional probability that the system, having reached interface $i$, subsequently goes on to reach interface $i+1$ before returning to the initial state. FFS provides a practical and efficient way to compute $\Phi$, and  $\ppi$ for each interface, thus allowing the computation of $k_{AB}$. The algorithm also generates  transition trajectories. While several variants of FFS exist \cite{FFS2,ffs,escobedorev}, we focus here on the direct FFS algorithm (DFFS) \cite{ffs,fFFS,escobedorev}.

The DFFS algorithm has two stages.  In the first stage, the flux $\Phi$ across the first interface $\lambda_0$ is computed by simulating a system in the initial state and monitoring the frequency with which the trajectory crosses $\lambda_0$ in the direction of increasing $\lambda$. When these crossings happen, the configuration of the system is stored; this simulation thus generates not only a measurement of $\Phi$ but also a collection of $N_0$ configurations corresponding to states of the system at the moments of crossing $\lambda_0$ 
\footnote{Note that if the system enters the final state during this run it should be returned to the initial state and re-equilibrated. It is also important to note that correct computation of $k_{AB}$ depends on good sampling of the configurations at $\lambda_0$: to avoid correlation between configurations arising from rapid recrossings of $\lambda_0$ it is often best to store configurations every $m$ crossings (with $m \approx 10$) rather than at every crossing -- see ref \cite{ffs} for further practical details.}. In the second stage of the algorithm, the probabilities $\ppi$ are computed in a step-wise fashion (see Figure \ref{fig:ffs_fig} for illustration). To compute $p_1$, one chooses configurations at random from the collection stored at $\lambda_0$  and uses them to initiate new "trial" trajectories, which are continued until they either reach $\lambda_1$ (``success'') or return to $\lambda_0$ (``failure''). For successful trajectories, the final configuration at $\lambda_1$ is stored in a new 
collection. After $M_0$ trial trajectories have been fired,  $p_1$ is computed by dividing the number of successes by $M_0$.  One then repeats the same  procedure, using the configurations at $\lambda_1$ as starting points for $M_1$ trajectories that are continued until they reach $\lambda_2$ or return to $\lambda_0$, and so on, until the final interface is reached and one has a complete set of estimated probabilities  $\ppi$. Transition trajectories from the initial to the final state can then be reconstructed from the set of successful trajectories between interfaces~\cite{FFS2,ffs}.

\subsection{Equalization of fluxes between interfaces}
Under the assumption that trajectories decorrelate between adjacent interfaces, analytic results for the computational efficiency of the DFFS method (and related methods) can be derived \cite{efficiency}. Even though these assumptions may not always be satisfied for many real problems, these analytical predictions still give a useful general guide to the performance of the method. In particular, by modelling the  number of successful trajectories from interface $i$  as a binomially distributed random variable with parameter $\ppi$, one can obtain predictions for  the computational cost, and the statistical error, associated with the computation of the rate constant $k_{AB}$ for given choices of the number $n$ of interfaces, the numbers $M_i$ of trial trajectories, the number $N_0$ of configurations at $\lambda_0$, and for given values of $\ppi$ \cite{efficiency}. For DFFS, the variance $\mathcal{V}$ in the estimated rate constant is given approximately by \cite{ffs,efficiency}
\begin{equation}\label{eq:var}
\mathcal{V} \approx \sum_{i=0}^{n-1}\frac{(1-\ppi)}{M_i \ppi}. 
\end{equation}
Borrero and Escobedo \cite{borrero,escobedorev} have shown that, for fixed $n$, $\{M_i\}$ and $P_B$, Eq.~(\ref{eq:var}) can be minimized by placing the interfaces such that  $M_i \ppi$ is the same for all interfaces -- i.e. the statistical error is smallest when the net flux of trajectories between successive interfaces is constant. Assuming, for simplicity, that one fires the same number of trajectories for each interface ($M_i = M=\text{const}$), one should  place the interfaces such that $\ppi$ is the same for all interfaces. Thus, interfaces should be  closer together in ``bottleneck'' regions of the phase space (note that here Borrero and Escobedo assume that, since the number of interfaces is fixed, the computational cost does not depend strongly on the interface placement, and does not need to be considered in the optimization). An alternative formulation of the constant flux rule, put forward by Borrero and Escobedo, states that the quantity 
\begin{equation}
f_i = \frac{\sum_{j=0}^{i-1} \log{p_j}} 
{\sum_{j=0}^{n-1} \log{p_j}}
\label{eq:finter}
\end{equation}
should be linear when plotted against the interface index $i$. To see this we note that if all the transition probabilities are equal,  $p_j=p=\text{const}$, then  
\begin{equation}
 f_i = \frac{\sum_{j=0}^{i-1} \log{p}} 
{\sum_{j=0}^{n-1} \log{p}} = \frac{i}{n}.
\label{eq:fconst}
\end{equation}
On can therefore measure the "quality" of a particular set of interfaces, either by directly asking whether the success probability $\ppi$ is the same across different interfaces, or by testing whether $f_i$ is linear when plotted against the interface number $i$. 

\subsection{Optimal transition probability}\label{sec:bof}

\begin{figure}[!ht]
\begin{center}
{\rotatebox{0}{{\includegraphics[scale=0.6,clip=true]{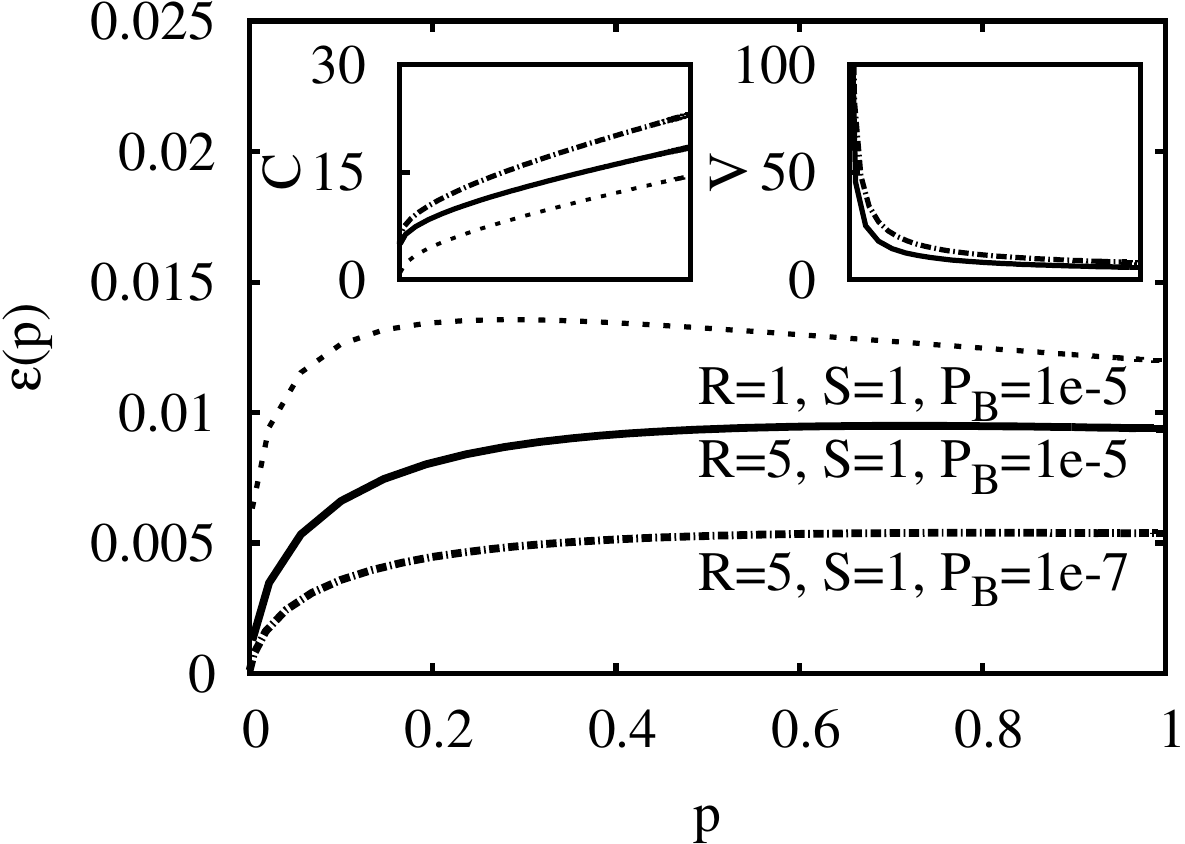}}}}
\caption{Theoretical prediction (Eq.(\ref{eq:eff})) for the efficiency $\mathcal{E}$ of a hypothetical rare event problem with $M=200$ and $N_0=100$, plotted as a function of $p$ for several values of $P_B$ and of the cost parameters $R$ and $S$ (see text and Appendix \ref{sec:ofc}). The insets show the predicted computational cost $\mathcal{C}$  (Eq.(\ref{eq:costpb})) and variance in the rate constant $\mathcal{V}$ (Eq.(\ref{eq:varpb}), which follows from Eq.(\ref{eq:var})).}
 \label{fig:cvem}
\end{center}
\end{figure}

Borrero and Escobedo's work shows that  for $n$ interfaces, the optimal positioning is such that 
$\ppi = p = P_B^{1/n}$ (this follows from Eq.~(\ref{eq:ppi})). However, if we are to place interfaces optimally, on-the-fly, we also wish to optimize the  number $n$ of interfaces. This is equivalent to optimizing the crossing probability $p$, under the constraint that $n=\log{P_B}/\log{p}$. Here we compute the optimal value of $p$, and we find that  the efficiency is rather insensitive to $p$ over a broad range of parameter values.

In optimizing the efficiency with respect to $p$ we need to consider both  computational cost and statistical error, since we expect both to depend on $p$. Following Ref. \cite{efficiency}, we define the  computational efficiency of a DFFS calculation as 
\begin{equation}\label{eq:eff1}
\mathcal{E} = \frac{1}{\mathcal{C}\mathcal{V}}
\end{equation}
where $\mathcal{V}$ measures the statistical error, as in Eq.(\ref{eq:var}), and $\mathcal{C}$ is the computational cost of the calculation. An approximate expression for $\mathcal{C}$ in terms of $p$ is  given in Appendix \ref{sec:ofc}, where we also rewrite  Eq.(\ref{eq:var}) for  $\mathcal{V}$ in terms of $p$ and $P_B$. Combining these expressions allows us to 
write an approximate analytical expression for the efficiency $\mathcal{E}$, as a function of $p$. This expression is given in Appendix \ref{sec:ofc}: it depends only on $p$, $P_B$, $M$ and two constants $R$ and $S$ which measure the cost of generating a single configuration at $\lambda_0$, and the cost of a trajectory from $\lambda_A$ to $\lambda_B$ (see Appendix \ref{sec:ofc}). Figure \ref{fig:cvem} shows  the function $\mathcal{E}(p)$, plotted for a hypothetical rare event problem in which $M=200$ and $N_0=100$, for several values of $P_B$, $R$ and $S$. $\mathcal{E}(p)$ is non-monotonic, with a peak at the optimal value of $p$. This non-monotonicity 
arises from contrasting trends in the computational cost and the statistical error (see insets to Figure \ref{fig:cvem}): while the cost increases with $p$ (because increasing $p$ implies more interfaces, and thus more trajectories), the statistical error decreases with $p$ (since more interfaces implies more accurate measurement of $k_{AB}$). 
 
Encouragingly, for all the parameter combinations tested, the peak in $\mathcal{E}(p)$ in Figure \ref{fig:cvem} is rather broad, suggesting that one may achieve high computational efficiency without needing to control the interface crossing probability too precisely. For low values of $p$ (less than about 0.2) the efficiency does, however, decrease drastically. Thus our calculations suggest that placing the interfaces such that the success probability is larger than about 0.3 should generally result in high computational efficiency. However, there is of course an upper limit on the value of $p$ that is sensible. For very high crossing probabilities, the interfaces become very close together and trajectories between successive interfaces become correlated -- a factor that is not taken into account in our theoretical analysis.  This is likely to compromise efficiency because correlated measurements at closely spaced interfaces incur computational overheads but provide little extra information.  Taking this 
into 
account, choosing a value of $p$ in the range $0.3 < p < 0.7$ appears to be a sensible rule of thumb for most rare event simulation problems. 

\section{Algorithms for automatic on-the-fly interface placement}\label{sec:ipm}

We now present two algorithms which automatically position the interfaces during a DFFS simulation so as to achieve a  desired value of the success probability $p$.
At the beginning of the simulation, the positions of the interfaces are not defined: the user specifies only the boundaries of the initial and final states ($\lambda_A$ and $\lambda_B$) and the desired value (or range) for $p$, as well as a minimal distance between interfaces so as to avoid correlations.  Starting from $\lambda_A \equiv \lambda_0$, the algorithms place interfaces  on-the-fly -- i.e. first the optimal value of $\lambda_1$ is determined, then trajectories are fired to $\lambda_1$, then $\lambda_2$ is optimized, then trajectories are fired to $\lambda_2$, etc. In these algorithms, the number $n$ of interfaces adapts automatically to the choice of $p$.

In order to place $\lambda_{i+1}$ optimally, given that the simulation has arrived at $\lambda_i$, one needs to make an estimate of how the transition probability $p_i$ depends on $\lambda_{i+1}$. Both algorithms achieve this by firing a small number of ``exploratory'' trajectories from $\lambda_i$; the difference between the two algorithms lies in the way that the information from these exploratory trajectories is used. 

\subsection{Trial interface method}

\begin{figure}
\begin{center}
{\rotatebox{0}{{\includegraphics[scale=0.3,clip=true]{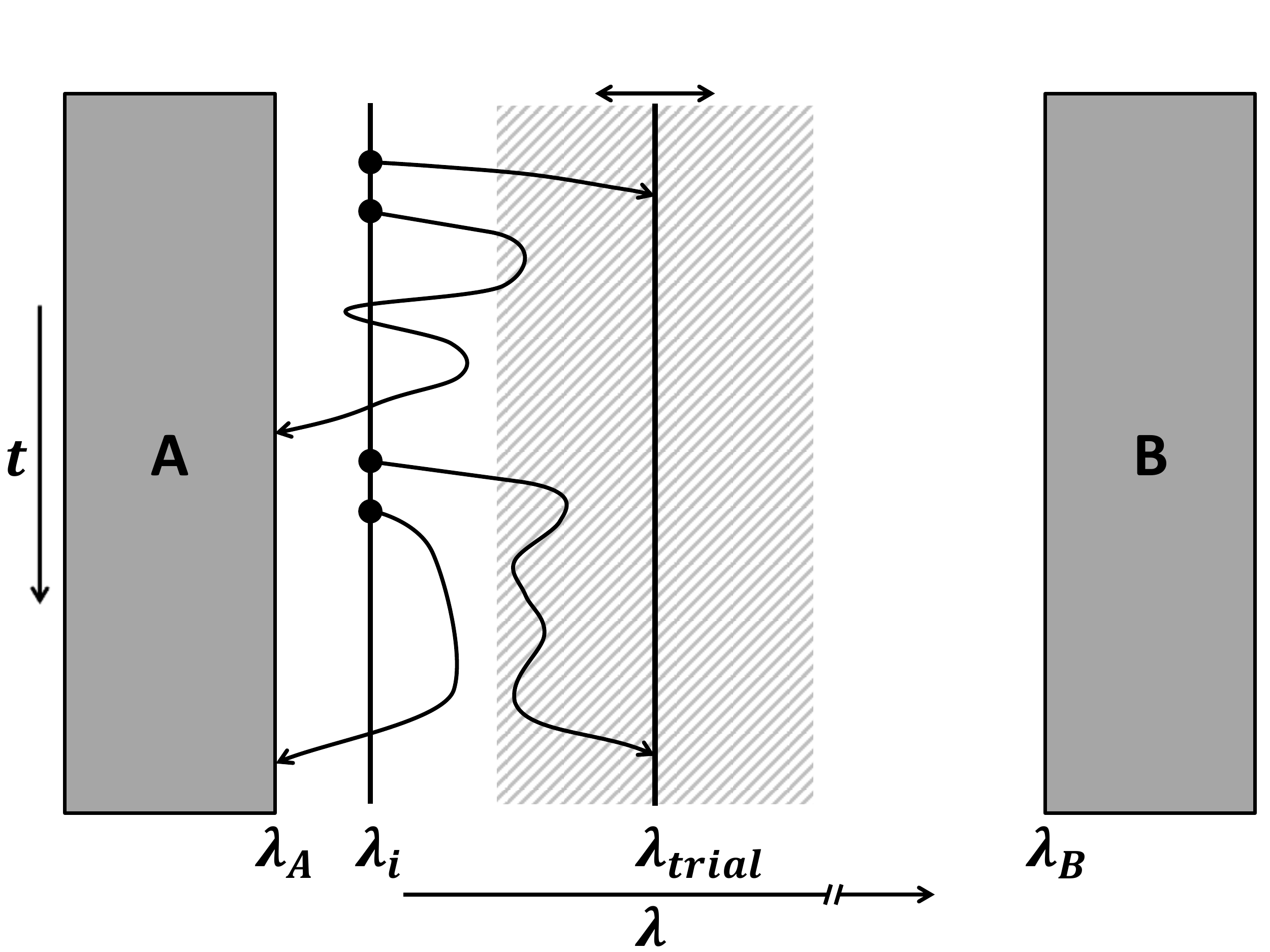}}}}
\caption{Schematic illustration of the trial interface method. The probability of reaching the trial interface $\lambda_\text{trial}$
  is estimated by firing a small number of trial runs from the current interface $\lambda_i$. The position of the trial interface is accepted if the estimated probability $p_{\text{est}}$ is in the range 
  $p_{\text{est}}\in[p_{\text{min}}, p_{\text{max}}]$. Otherwise, the trial interface is shifted according to Eq.(\ref{eq:trials}) \label{fig:auto_virtual}}
\end{center}
\end{figure}

In the ``trial interface'' method we position interfaces such that the transition probability $p$ lies within user-defined acceptable bounds  $p\in[p_{\text{min}}, p_{\text{max}}]$.
To achieve this, we choose a trial position $\lambda_{\text{trial}}$ for the next interface, obtain an estimated transition probability $p_{\text{est}}$  for this trial interface, and then, based on this information, shift the trial interface until  $p_{\text{est}}$ lies within the range $p_{\text{est}}\in[p_{\text{min}}, p_{\text{max}}]$.  Assuming that the DFFS simulation has reached interface $\lambda_i$, the algorithm proceeds as follows (see also Figure \ref{fig:auto_virtual}):

\begin{enumerate}
\item{Choose a trial position $\lambda_{\text{trial}}$ for the next interface $\lambda_{i+1}$, in the range $\lambda_i < \lambda_{\text{trial}} < \lambda_B$. This should be done in a way appropriate to the problem being studied; we typically set  $\lambda_{\text{trial}} = \lambda_i + b\times (\lambda_B-\lambda_A)$, where $0.01<b<0.1$, but one could also use for example $\lambda_{\text{trial}} = \lambda_i + (\lambda_i-\lambda_{i-1})$. }.
\item{Using as starting points the configurations stored at $\lambda_i$, fire $M_{\text{trial}}$ trajectories, which are continued until they reach $\lambda_{\text{trial}}$ or $\lambda_A$. $M_{\text{trial}}$ should be significantly smaller than the typical number of $M$ trajectories per interface in the complete FFS simulation. }
\item{Compute $p_{\text{est}}=N_{\text{S}}/M_{\text{trial}}$ where $N_{\text{S}}$ is the number of trial trajectories that reached $\lambda_{\text{trial}}$.}
\item{If $p_\text{min} < p_{\text{est}} < p_\text{max}$, accept the trial interface. Otherwise, choose a new trial interface position according to 
\begin{equation}\label{eq:trials}
  \lambda_\text{trial, new} =
  \lambda_\text{trial, old} + 
  \lambda_{\text{step}} \Delta p
\end{equation}
where 
\begin{equation}
\Delta p = 
\begin{cases} (p_{\text{est}}-p_{\text{max}}) & 
    \text{if } p_{\text{est}} > p_{\text{max}} \\
    (p_{\text{est}}-p_{\text{min}}) & 
    \text{if } p_{\text{est}} < p_{\text{min}}
  \end{cases}
 \end{equation}
and fire trial trajectories to obtain a new $p_{\text{est}}$ for this trial interface. Repeat this procedure until $p_{\text{est}}$ lies within the desired range. If the resulting value $\lambda_\text{trial, new} <\lambda_i+ d_{\text{min}}$, set  $\lambda_\text{trial} = \lambda_i + d_{\text{min}}$, where $d_{\text{min}}$ is the user-defined minimal acceptable distance between interfaces. If $\lambda_\text{trial, new} > \lambda_B$ set $\lambda_\text{trial} = \lambda_B$.}
\item{Set $\lambda_{i+1} = \lambda_{\text{trial}}$.}
\item{Continue with the DFFS simulation -- i.e.  fire $M$ trajectories to $\lambda_{i+1}$ to compute $p_i$ and obtain a new collection of configurations at $\lambda_{i+1}$, as in the standard DFFS procedure. Any trajectories previously fired to this interface during step 5 can be included in the estimate of $p_i$.}
\end{enumerate}

In this algorithm,  in addition to $p_{\text{min}}$ and $p_{\text{max}}$, the user-defined parameters are the stepwidth $\lambda_{\text{step}}$ (which determines how far the interface is shifted in each adjustment step), $M_{\text{trial}}$, the number of trial trajectories used to obtain $p_{\text{est}}$ (typically  $M_{\text{trial}}\approx 15$), and $d_{\text{min}}$, the minimal acceptable spacing between interfaces. This latter parameter is introduced to avoid excessive correlation between the sampling at successive interfaces, even if $p_{\text{min}}$ is chosen to be small. The choice of $d_{\text{min}}$ depends on the choice of order parameter and the dynamics of the system being studied. For example, if the order parameter is discrete, $d_{\text{min}}$ should be at least one. In continuous systems, it should  prevent the system from being able to cross several interfaces in a single timestep, and should be larger for systems whose dynamics is slow to decorrelate.

The choice of interface shifting rule (point 4 in the algorithm described above) is not unique. We expect this rule to work well for systems with steep energy barriers, where one needs the initial interfaces to be closely spaced. However, for systems with  flatter barriers,   one might prefer to use a bisectional scheme, in which  the trial interface is  initially placed midway between $\lambda_i$ and $\lambda_B$, and is then shifted forwards or backwards by bisecting the space between itself and either $\lambda_i$ or $\lambda_B$.

The trial interface method is conceptually simple and can be implemented with only very minor modifications to  an existing DFFS simulation code. The method also has the advantages that estimated transition probabilities for several possible trial interface positions can be computed in parallel on separate processors, and that any trial trajectories fired to interfaces that are eventually accepted can be reused in the final calculation of $p_i$. The method does, however, have the potential drawback that it relies on a reasonably good first estimate of $\lambda_{\text{trial}}$: if this first estimate is very poor, the algorithm may take many iterations to find an acceptable interface position. This problem is avoided in our second approach, the ``exploring scouts'' method. 

\subsection{Exploring scouts method}

\begin{figure}
\begin{center}
  {\rotatebox{0}{{\includegraphics[scale=0.3,clip=true]{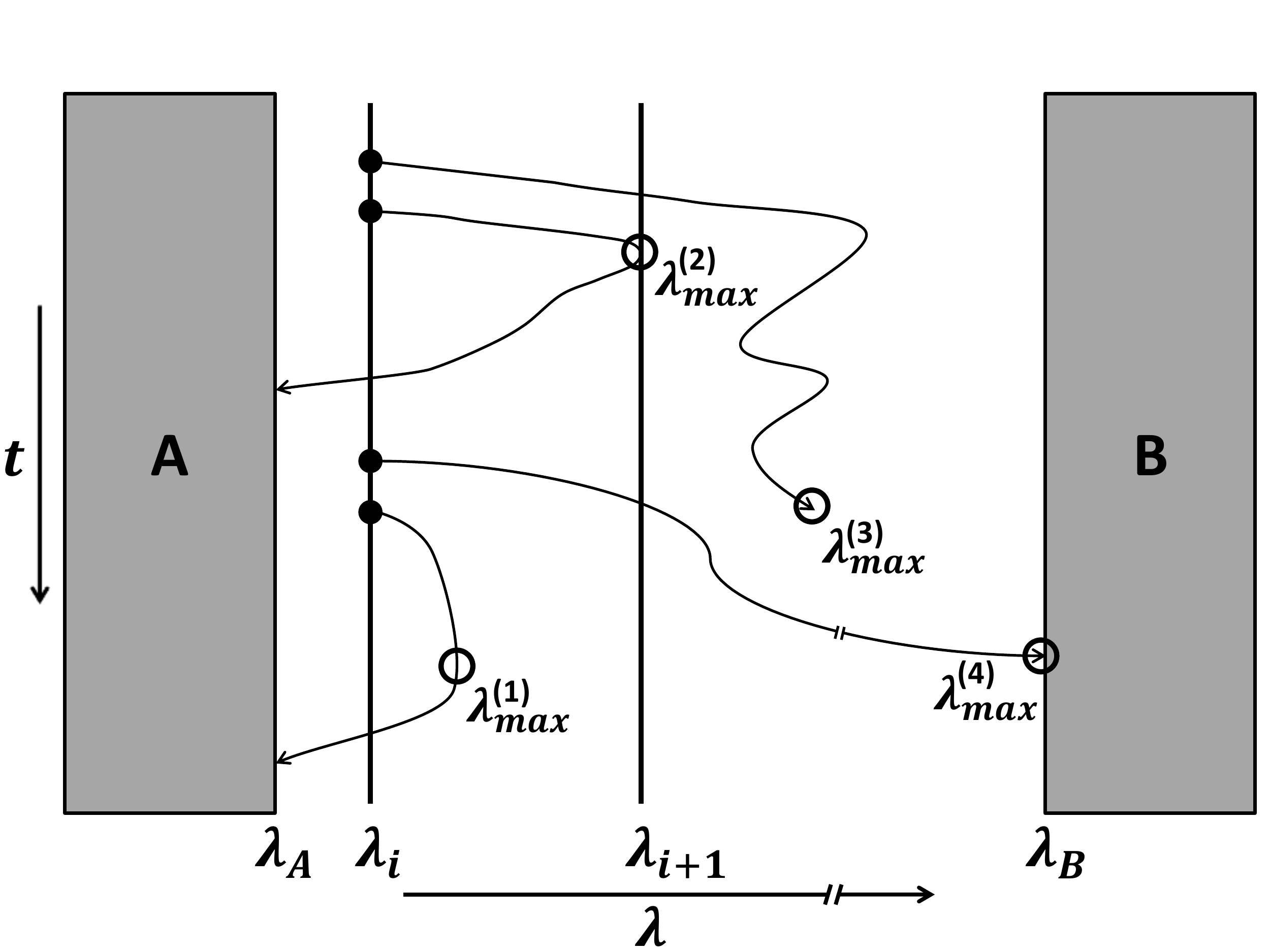}}}}
  \caption{Schematic illustration of the exploring scouts method. A pre-defined number of trial trajectories are launched from the current interface $\lambda_i$. These trajectories continue until they reach the initial state A or the final
    state B, or until the maximum number of steps is reached. The maximum values of $\lambda$ reached by the trial trajectories are then used to determine the position $\lambda_{i+1}$ of the next interface. \label{fig:auto_scout}}
\end{center}
\end{figure}

In the "exploring scouts" method, we again fire $M_{\text{trial}}$ trial trajectories from interface $\lambda_i$, but this time without defining a trial interface position.  In this method, illustrated in Figure \ref{fig:auto_scout}, the trial trajectories are continued until they reach either $\lambda_B$ or $\lambda_A$, or until a user-defined maximum number of steps is exceeded. The maximum value of $\lambda$ achieved by each trial trajectory is monitored, and the distribution of these values is used to position the next interface such that the success probability is close to a user-defined desired value $p_{\text{des}}$. The exploring scouts algorithm proceeds as follows:
\begin{enumerate}
\item{Fire $M_{\text{trial}}$ trial trajectories from interface $\lambda_i$, starting from the configurations generated  by the DFFS algorithm. Continue each trajectory until it either reaches  $\lambda_B$ or $\lambda_A$, or exceeds $m_{\text{max}}$ steps. Record the maximum value of $\lambda$ achieved in each trial trajectory.}
\item{Generate a ranked list of maximum $\lambda$ values for all trial trajectories -- i.e. assign each trajectory an index $k$ in the range $0 < k < M_{\text{trial}}$, such that $\lambda_{\text{max}}^{(k)} < \lambda_{\text{max}}^{(k+1)}$.}
\item{Compute $k_{\text{des}} = \lfloor M_{\text{trial}}(1-p_\text{des}) \rfloor$ 
and set the position of the next interface $\lambda_{i+1} = \lambda_{\text{max}}^{(k_{\text{des}})}$. If the resulting value $\lambda_{i+1} < \lambda_i+ d_{\text{min}}$, set  $\lambda_{i+1} = \lambda_i + d_{\text{min}}$ (where $d_{\text{min}}$ is the minimal acceptable spacing between interfaces as in the trial interface method).}
\item{Continue with the DFFS simulation -- i.e.  fire $M$ trajectories to $\lambda_{i+1}$ to compute $p_i$ and obtain a new collection of configurations at $\lambda_{i+1}$, as in the standard DFFS procedure. }
\end{enumerate}

This algorithm works because the trial trajectories, or ``exploring scouts'', supply information on the probability of reaching a particular value of $\lambda$, for all $\lambda$ in the range $\lambda_i \to \lambda_B$. For entry $k$ 
in our ranked list,  
$k$ exploring scouts failed to reach $\lambda_{\text{max}}^{(k)}$ and $M_{\text{trial}}-k$ scouts reached $\lambda_{\text{max}}^{(k)}$ or beyond (note $k$ runs from zero to $M_{\text{trial}}-1$). 
The transition probability for an interface placed at $\lambda_{\text{max}}^{(k)}$ would therefore be approximately $(M_{\text{trial}}-k)/M_{\text{trial}}$. We can obtain a next interface position $\lambda_{i+1}$ corresponding approximately to our desired transition probability $p_{\text{des}}$ simply by picking the $\lfloor M_{\text{trial}}(1-p_\text{des}) \rfloor$-th entry in our list of maximal $\lambda$ values. More precise versions of this algorithm are of course possible (e.g. interpolating between  $\lambda_{\text{max}}^{(k)}$ values in our list). However, because the efficiency is in general not very sensitive to the precise value of $p$, we do not find these to be necessary.

The user-defined parameters for this method are the target probability $p_{\text{des}}$, the  number $M_{\text{trial}}$ of exploring
scouts, the minimal interface spacing $d_{\text{min}}$ and the limit $m_{\text{max}}$ on the number of simulation steps per trial trajectory.  If $m_{\text{max}}$ is set too low, the algorithm will fail to explore regions of larger $\lambda$, and may tend to place the interfaces too close together (i.e. the true $p$ will be smaller than $p_{\text{des}}$). 
Choosing a  large value of $m_{\text{max}}$ will, however make the algorithm more computationally expensive.  

The exploring scouts method has the advantage that one knows {\em{a priori}} how many trial trajectories will be required to set the next interface position -- this may be important in parallelized FFS applications. Furthermore, the number of user-defined parameters is fewer than in the trial interface method. The exploring scouts method requires slightly more modifications to an existing standard DFFS code than the trial interface method, since one needs to track the maximal values of $\lambda$ for the trial trajectories, but it is nevertheless rather simple to implement. 

\section{Examples}\label{sec:simex}

We now demonstrate  our interface placement methods for two test problems. First, we study the toy problem of a single particle undergoing Langevin dynamics in a one-dimensional potential; this also provides an opportunity to test the predictions for the computational efficiency made in section \ref{sec:bof}. Next, we demonstrate the utility of the methods for the much more computationally demanding example of crystal nucleation in a system of  particles interacting {\em{via}} a Yukawa potential.

\subsection{A single particle in a one-dimensional potential}\label{sec:ex1D}

We first consider a single particle moving in one dimension, in a potential with two minima, defined by $V(x)=(h/2)\left[1-\cos{(\pi x)}\right]$ for $x$ in the range $[-1,3]$. The height of the potential barrier, at $x=1$, is $h=12k_BT$. 
The particle, which is initially placed in the region $-0.2<x<0.2$, undergoes underdamped Langevin dynamics. We set $k_BT=1$, $m=1$, $dt=0.001$ and the friction coefficient $\gamma=1$; with these parameters the crossover between ballistic and diffusive motion occurs on a timescale of about 1000 time steps or a dimensionless distance of 1. Our reaction coordinate $\lambda$ is taken to be the position $x$ of the particle and the borders of the initial and final states are defined by $\lambda_A=0.2$ and $\lambda_B=2$.

We carry out DFFS simulations for this problem, using both the trial interface method and the exploring scouts method. For both methods, we set  $M_{\text{trial}}=100$, $M=1000$, $N_0=3000$ and $d_{\text{min}}=0.01$. In the trial interface method, we set $p_{\text{min}}=0.4$, $p_{\text{max}}=0.6$ and the initial trial position for interface $\lambda_{i+1}$ is chosen to be $\lambda_i+0.1(\lambda_B-\lambda_i)$. In the exploring scouts method, we set $p_{\text{des}}=0.5$ and $m_{\text{max}}=10^5$.

\begin{figure}
\begin{center}
{\rotatebox{0}{{\includegraphics[scale=0.55,clip=true]{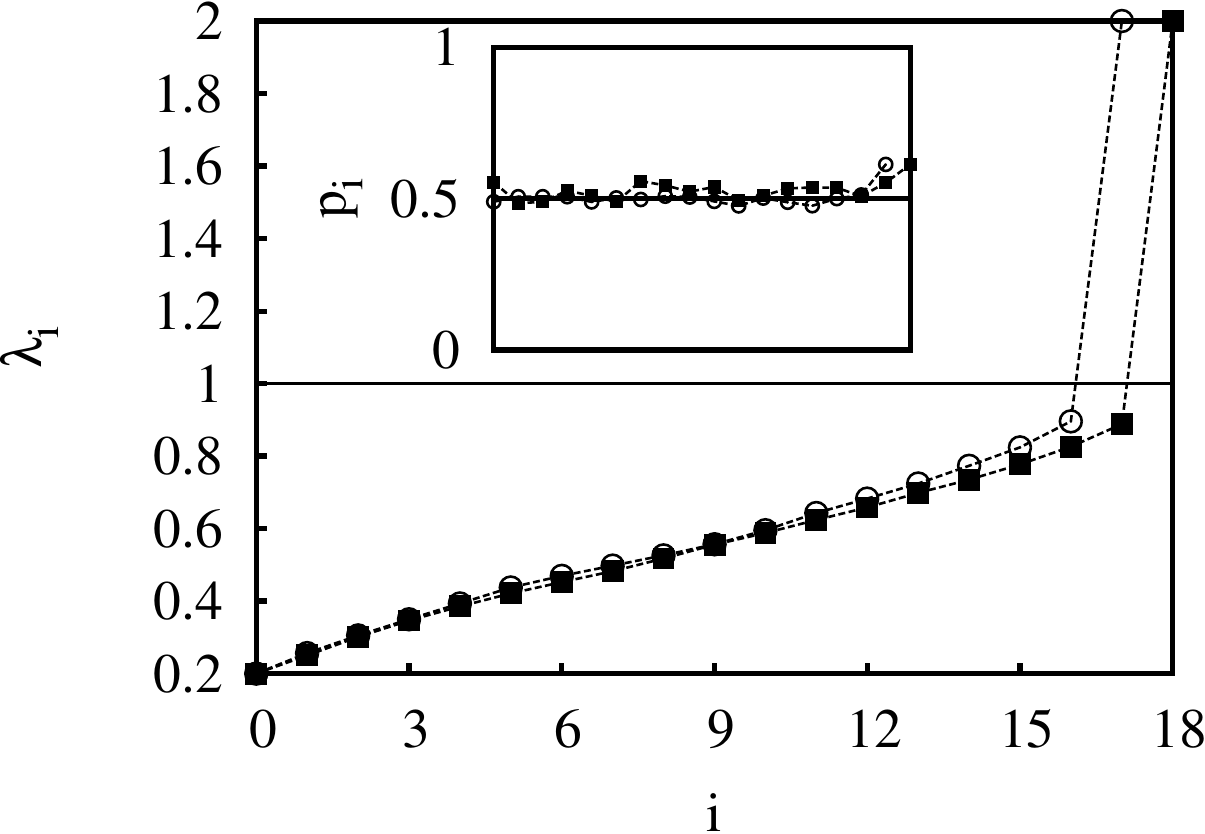}}}}
\caption{Single particle test case: positions $\lambda_i$ of the interfaces as a function of interface index $i$, for the trial interface method (squares) and the exploring scouts method (circles). The maximum of the potential barrier is at $\lambda=1.0$, shown by the solid horizontal line. The inset shows the success probabilities $p_i$, plotted as a function of interface index $i$. \label{fig:splp}}
\end{center}
\end{figure}
\begin{figure}
\begin{center}
{\rotatebox{0}{{\includegraphics[scale=0.55,clip=true]{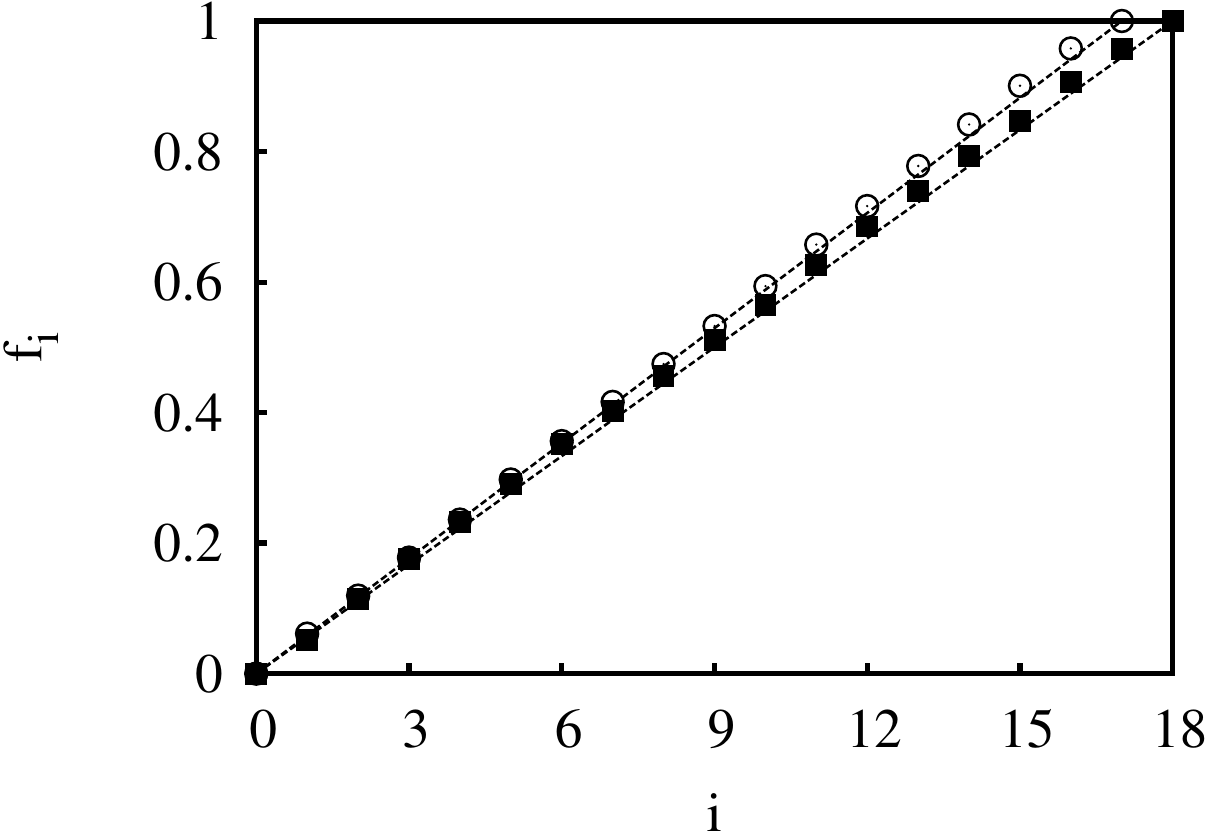}}}}
\caption{Single particle test case: $f_i$, as defined in Eq. (\ref{eq:finter}), as a function of interface index $i$, for the trial interface method (squares) and the exploring scouts method (circles). The dashed lines show the optimal case where $f_i=i/n$ (see Eq. (\ref{eq:fconst})). Note, that the two methods give slightly different numbers of interfaces.
\label{fig:ftifes}}
\end{center}
\end{figure}

Figure \ref{fig:splp} shows the positions of the interfaces (main plot), and the resulting success probabilities $p_i$ (inset), for the trial interface and exploring scouts methods.  Both methods produce probabilities $p_i$ that are approximately uniform across the interfaces, as desired. This corresponds to a highly non-uniform interface spacing: in fact all the interfaces are located prior to the maximum of the potential barrier, with the final interface lying close to the maximum. Figure
\ref{fig:ftifes} shows the quantity  $f_i$, defined in Eq.(\ref{eq:finter}), for both methods. This is indeed close to linear, confirming that the interface placement is close to optimal. 
 
These methods allow us to choose the success probability $p$ and place interfaces accordingly. We can therefore use them to test the theoretical predictions made in section \ref{sec:bof} for the dependence of the computational efficiency on $p$.
To this end, we have used the exploring scouts method to carry out a series of DFFS simulations for the single particle test problem, with the transition probability $p$ varying between $0.05$ and $0.95$. The parameters of the method were as above, but with $M=3000$. In these simulations, we measured the computational cost (in simulation steps) and the statistical error in the computed rate constant, allowing us to compute the computational efficiency $\mathcal {E}$ as defined by Eq.(\ref{eq:eff1}).
\begin{figure}
\begin{center}
{\rotatebox{0}{{\includegraphics[scale=0.55,clip=true]{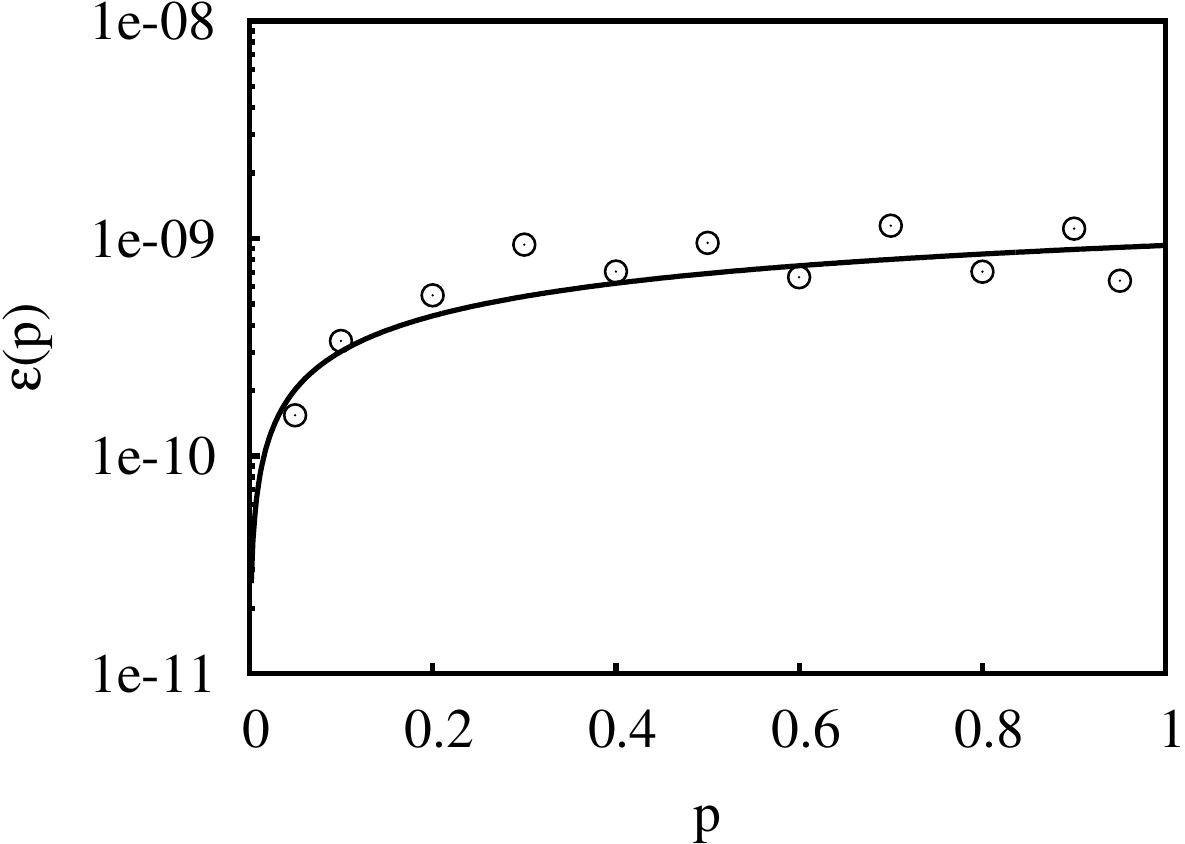}}}}
\caption{Single particle test case: computational efficiency $\mathcal {E}$ as defined by Eq.(\ref{eq:eff1}), as a function of the success probability $p$. The data points show results of simulations using the trial interface method (for parameters see main text). The solid line shows the theoretical prediction of Eq.~(\ref{eq:eff}) with $P_B=1.36\times 10^{-5}$, $R=1.60\times 10^7$ and $S=1.39\times 10^7$ (see main text). The number of interfaces placed by the algorithm varied between $3$ (for $p=0.05$) and $671$ (for $p=0.95$).
\label{fig:effsp}}
\end{center}
\end{figure}
Figure \ref{fig:effsp} shows the measured computational efficiency, as a function of the transition probability $p$, compared to the theoretical prediction. The latter was computed using Eq.(\ref{eq:eff}), with $P_B=1.36\times 10^{-5}$ (the result obtained from our simulations), $R=1.60\times 10^7$ and $S=1.39\times 10^7$ (both in simulation steps, and obtained by fitting the cost function \eqref{eq:costpb} to our simulation data). The simulations are in remarkably good agreement with our theoretical predictions, showing that the estimated optimal values of $p$ obtained from the theory are indeed valid, at least for this problem. Taking the error bars into account, the value of $P_B$ obtained in the our simulations is independent of $p$, justifying the use of  $p$ as a performance tuning parameter and showing that  the  FFS method remains valid regardless of the number of interfaces (which varies between 3 and 671 as $p$ is varied between $0.05$ and $0.95$).

\subsection{Crystallization of Yukawa particles}
We now move on to a much more challenging test problem: crystal nucleation in a system of particles interacting {\em{via}} a combined Yukawa and Weeks-Chandler-Andersen (WCA) potential \cite{weeks71a}, $U(r) = U_\text{Yukawa}(r) + U_\text{WCA}(r)$ with
\begin{equation}
 U_\text{Yukawa}(r) = \epsilon \frac{\exp(-\kappa(r/\sigma-1))}{r/\sigma}
\label{eq_yukawa}
\end{equation}
and
\begin{equation}
 U_\text{WCA}(r) =
\begin{cases}
 4 \left( \left( \frac{\sigma}{r} \right)^{12} - \left( \frac{\sigma}{r} \right)^{6} + \frac{1}{4} \right) & r < \sigma^\frac{1}{6} \\
 0 & \text{else}.
 \end{cases}
\label{eq_wca}
\end{equation}
The repulsive WCA potential is used to model the excluded volume of the particles (note that the energy scale for
the WCA potential is set to $k_BT=1$ in our simulation units). The Yukawa potential is a screened Coulomb potential, suitable for modeling charged particles whose electrostatic interactions are screened by surrounding ionic atmospheres. 
In this work, the parameters of the Yukawa potential are the value of the repulsive potential at contact $\epsilon=8$ (in units of $k_BT$) and the inverse screening length $\kappa=5$ (in terms of the hard-sphere diameter $\sigma$). 
Despite important previous advances \cite{yukawa,sanz2007}, the mechanism by which crystal nucleation happens in screened Coulomb systems remains an open question, to which FFS simulations can contribute by providing both nucleation rates and transition paths \cite{yukawa}. However, because the Yukawa interaction requires a larger cutoff radius
than the more widely studied Lennard-Jones interaction, simulations of Yukawa particles are computationally expensive (especially for low salt conditions), which means that the number of trial trajectories which can be performed in an FFS simulation is limited. This makes setting up standard FFS simulations difficult, particularly under interesting conditions, e.g. close to coexistence where the transition rate is expected to be low~\cite{azhar}. Under these conditions, manual placing of the
interfaces can easily lead to conditions where no FFS trial trajectories succeed in reaching the next interface. For such systems, automatic, optimal interface placement has the potential greatly to improve the feasibility and computational efficiency of FFS simulations. 

We performed molecular dynamics (MD) simulations of $4096$ WCA-Yukawa particles in a cubic box with periodic boundary conditions in the NPT ensemble at constant pressure $P=38$ (LJ units) with a Langevin thermostat using the software package ESPResSo \cite{espresso} in combination with DFFS, implemented in our rare event sampling framework FRESHS~\cite{freshs}. Note that FFS requires stochastic dynamics: here this is provided by the Langevin thermostat. The system is initially prepared in the liquid phase, which is undercooled (and therefore metastable). We are interested in the transition to the stable FCC crystal phase. Our order parameter $\lambda$ is the  size of the largest cluster of solid particles, where particles are classified as solid or liquid based on the local $q_6$ order parameter, as used in previous work \cite{q6}. The boundaries of the initial and final states were fixed such that the system is in the initial state if less than $0.5\%$ of the 
particles are in the largest solid cluster and in the final state 
if more than 90\% of the system's particles are in the largest solid cluster. This corresponds to $\lambda_A=15$ and $\lambda_B=3700$.

In our DFFS simulations, we compared three methods for interface placement: (i) placing the interfaces manually {\em{via}} a logarithmic scheme, (ii) the  trial interface method  and (iii) the exploring scouts method. All our DFFS simulations used $N_0=80$ configurations at the first interface and $M=50$ trial runs per interface. Here, we discuss only the performance of the interface placement methods; the nucleation rates and pathways  generated in the simulations will be presented elsewhere~\cite{kratzeryuk}.

\begin{figure}
\begin{center}
{\rotatebox{0}{{\includegraphics[scale=0.55,clip=true]{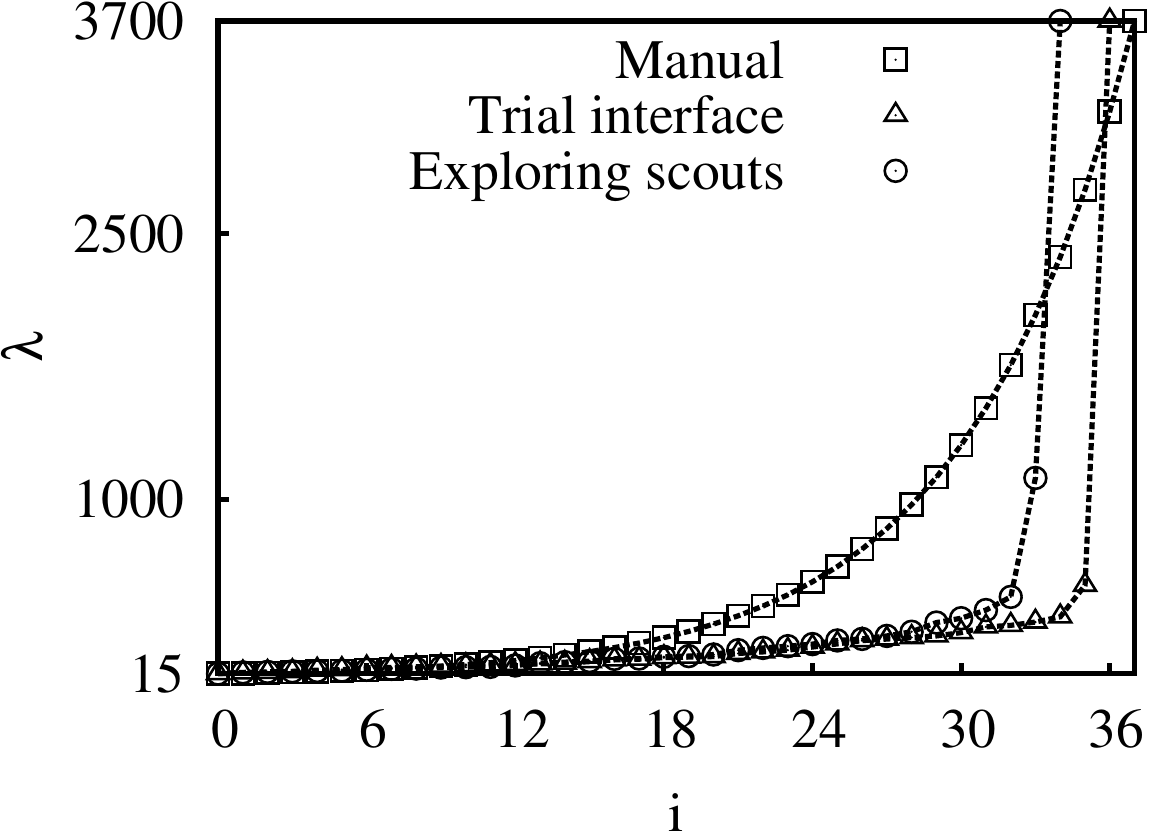}}}}
\caption{Yukawa test case: interface positions $\lambda_i$ (plotted as a function of interface index $i$) generated by the manual interface placement (open squares), the trial interface method (open triangles) and the exploring scouts method (circles).
  \label{fig:p_vgl}}
\end{center}
\end{figure}

\begin{figure}
\begin{center}
\makebox[20pt][l]{(a)}{\rotatebox{0}{{\includegraphics[scale=0.55,clip=true]{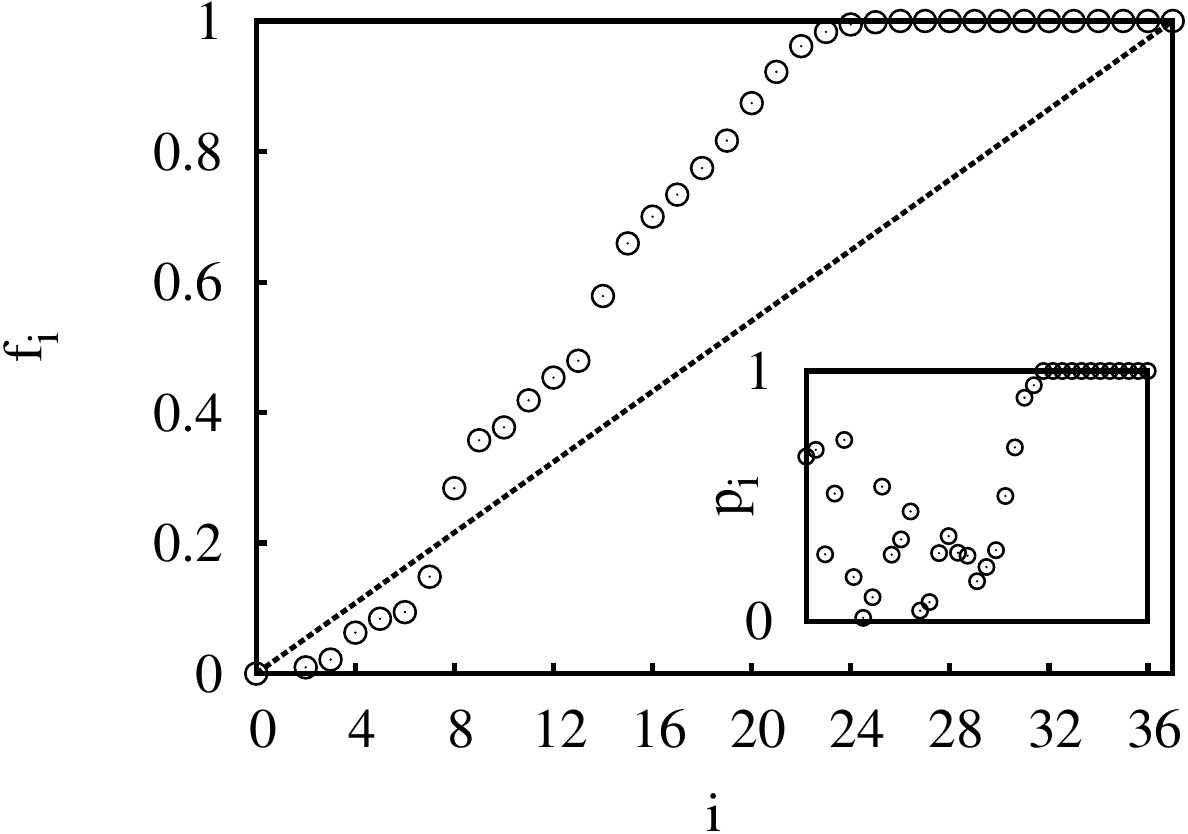}}}}
\makebox[20pt][l]{(b)}{\rotatebox{0}{{\includegraphics[scale=0.55,clip=true]{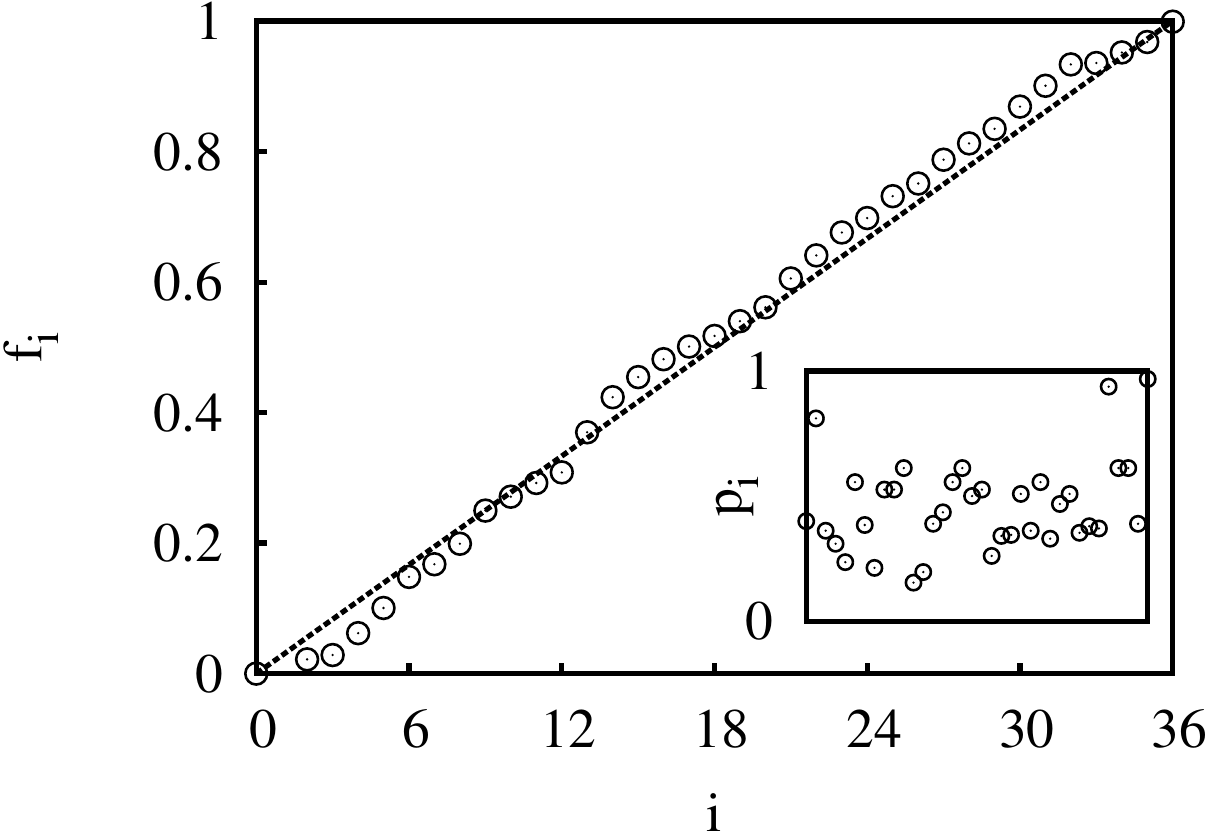}}}}
\makebox[20pt][l]{(c)}{\rotatebox{0}{{\includegraphics[scale=0.55,clip=true]{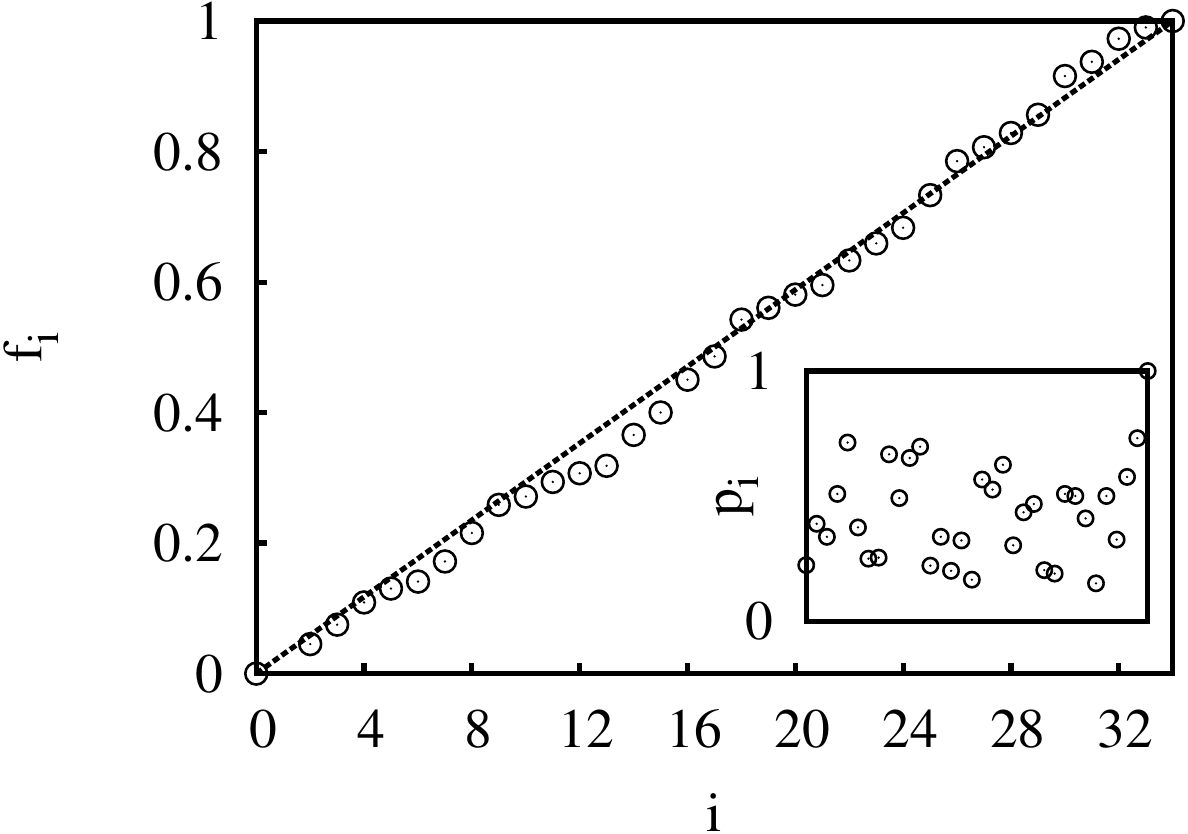}}}}
\caption{Yukawa test case: optimization criteria for the interface sets generated manually (a), with the trial interface method (b) and with the exploring scouts method (c) (for parameter values see main text). The main plots show the function $f_i$ of Eq. (\ref{eq:finter}), plotted against the interface index $i$, from our simulations (symbols) -- the dashed lines show the optimal case, Eq. (\ref{eq:fconst}). The insets show the success probabilities  $p_i$ plotted against the interface index $i$.   \label{fig:yuk_all}}
\end{center}
\end{figure}

We first discuss  the manual interface placement. For nucleation problems, where simulations are computationally very expensive, manual interface placement in FFS is very challenging. Our problem has a steep free energy barrier and so placing interfaces evenly between $\lambda_A$ and $\lambda_B$ results in very poor success rates for early interfaces. In fact, for our problem, we did not obtain any successes for early interfaces even with  $100$ evenly spaced interfaces. Therefore, as a ``best possible'' manual choice, based on this prior knowledge, we placed  $36$ interfaces logarithmically between  $\lambda_A$ and $\lambda_B$, with closer spacing between the early interfaces. Even with this rather well-informed choice of interfaces, 
Figure \ref{fig:yuk_all}(a) shows that we obtain success probabilities that are far from equal across interfaces (inset). Indeed, many of the $p_i$ values are very low: this is because the free energy landscape contains unforseen bottleneck regions, in which too few interfaces were placed. Because the success probabilities are low in these bottleneck regions, much computational effort will be wasted on failed trajectories. Another problem is also apparent: for later interfaces, the transition probabilities are extremely high (close to 1). In this region of the free energy landscape, the crystal grows spontaneously: the placement of unnecessary interfaces implies extra computational overhead in storing configurations, etc. The fact that the manual interface placement is far from optimal is also apparent in the highly non-linear form of the function $f_i$ when plotted against the interface index $i$ (main plot in Fig. \ref{fig:yuk_all}(a)). 

We note that a commonly used approach to manual interface placement in FFS is to start with some initial guess, then  if one obtains no successes for a given interface, shift it to a lower $\lambda$ value and continue the FFS simulation. If not done carefully, this can actually bias the resulting computation of the rate constant $k_{AB}$ towards higher values, since for interfaces at which  by
chance one obtains a large number of successes, one makes no change, but for interfaces where by chance one obtains few successes one shifts the interface. If such a shifting approach is used, bias can only avoided by repeating the entire FFS simulation {\em{a posteriori}} - i.e. after the interface positions have been fixed.

Figures \ref{fig:p_vgl} and \ref{fig:yuk_all} also show the results of the automatic interface placement methods. For both methods, we set  $d_{\text{min}}=3$  and $M_{\text{trial}}=8$. The value of $d_{\text{min}}$ was chosen to prevent the system from crossing several interfaces in one MD timestep (simultaneous attachment of 3 particles in one step is unlikely) and to avoid correlation between trajectories at successive interfaces. For the trial interface method, we used $p_{\text{min}}=0.3$ and $p_{\text{max}}=0.6$ and the initial trial position for $\lambda_{i+1}$ was set at $\lambda_i+0.1(\lambda_B-\lambda_A)$. For the exploring scouts method, we used $p_{\text{des}}=0.45$ and $m_{\text{max}}=10000$ timesteps. Figure \ref{fig:p_vgl} shows that both these methods produce similar interface numbers and positions, which are very different from those of our manual interface placement. The automatic methods position the interfaces much closer to the A state: in fact there are no interfaces at 
all for $\lambda$ values greater than $1120$. This suggests that the free energy barrier to nucleation is located 
closer to $\lambda_A$ than to $\lambda_B$ 
-- once the system has passed the barrier, the transition probability is always greater than the target value and thus no further interfaces are necessary.  However, without {\em{a priori}} knowledge, there would be no way to guess this when placing the interfaces manually. Figure \ref{fig:yuk_all} (b) and (c) show that indeed both automatic interface placement methods perform well according to our optimization criteria: the success probabilities $p_i$ are much more uniform, with no very low $p_i$ values (insets). The functions $f_i$ are also much more linear for the automatic interface placement methods than for the manual interface placement (main plots). 

\begin{center}
\begin{table}
\begin{tabular}{|c|c|}\hline
Method & $k_{AB}$ \\ \hline
Manual placing & $1\times 10^{-14\pm2}$ \\
Trial interface & $6\times 10^{-14\pm1}$ \\
Exploring scouts & $2\times 10^{-14\pm1}$ \\ \hline
\end{tabular}
\caption{Yukawa test case: rate constant $k_{AB}$ (in $\sigma^{-3} \tau^{-1}$ with the simulation time unit $\tau$) for DFFS simulations using the manual interface placement, trial interface method and exploring scouts method. For parameter values, see main text. The error bars in $k_{AB}$ were determined by repeated independent simulations.}
\label{tb:simdet}
\end{table}
\end{center}

\begin{center}
\begin{table}
\begin{tabular}{|c|c|c|c|}\hline
Method & Cost $\mathcal{C}$ & Variance $\mathcal{V}$ & Efficiency $\mathcal{E}$ \\ \hline
Manual placing & $7\times 10^6$ & $6648$ & $10^{-11}$\\
Trial interface & $4\times 10^6$ & $188$ & $10^{-9}$ \\
Exploring scouts & $3\times 10^6$ & $251$ & $10^{-9}$ \\ \hline
\end{tabular}
\caption{Yukawa test case: computational cost, variance in the rate constant and resulting computational efficiency, for DFFS simulations using the manual interface placement, trial interface method and exploring scouts method. For parameter values, see main text. The cost was measured in simulation timesteps including the cost of exploratory trial runs for the automatic interface placement methods. The variance in the rate constant was estimated  using Eq.~(\ref{eq:var}), using simulation data for the $p_i$ values.}
\label{tb:simdet2}
\end{table}
\end{center}

An obvious advantage to using the automatic interface placement methods is that setting up a DFFS simulation becomes very much easier and less time-consuming than using manual interface placement. In addition, the resulting DFFS calculations are more efficient with the optimized interface sets. Table  \ref{tb:simdet} shows that the rate constants computed using the three interface placement methods are equivalent, but the error bars (computed by repeated FFS calculations) are larger for the manual interface placement. Moreover, as shown in Table \ref{tb:simdet2}, the computational cost of the FFS calculation, measured in simulation steps, was about a factor of 2 lower for the automatically placed interfaces than for those that were placed manually. Had we not used our prior knowledge to place the manual interface set logarithmically, this factor would have been even greater. For this problem, the exploring scouts method required about $25\%$ fewer simulation steps than the trial interface method. Table 
\ref{tb:simdet2} also shows estimates for the statistical errror in the  rate constant (computed using Eq.  (\ref{eq:var})), and the resulting computational efficiency. The estimated computational efficiency is two orders of magnitude higher using the automatic interface placement methods, compared to the manual interface set.

\section{Discussion}\label{sec:dis}

The efficiency of interface-based rare event simulation methods such as FFS is strongly dependent on the locations of the interfaces. Without {\em{a priori}} information,  manual interface placement is a ``hit and miss'' task, that, for computationally intensive systems, often involves a large amount of user effort and results in non-optimal interface sets, for which the FFS calculations may be inefficient. In this paper, we have presented two methods for automatically placing interfaces on-the-fly in DFFS simulations, so that the user need only choose the order parameter, the definitions of the initial and final states, and the target transition probability (or its range). Building on previous work by Borrero and Escobedo, we have analysed theoretically how the computational efficiency depends on the interface transition probability $p$, providing an analytical expression for the optimal value of $p$ for 
a given  total transition probability $P_B$.  We have further shown that in fact this optimum is broad and not very sensitive to $P_B$, so that for most problems target success probabilities in the range $0.3-0.7$ are likely to produce satisfactory results. The lower bound of this range is set by the fact that efficiency decreases strongly when the success probability becomes too low. The upper bound is determined by the fact that trajectories will be highly correlated at successive interfaces if they are too close, meaning that little extra information is gained. 

Our two methods for automatic interface placement both work by firing a small number of trial trajectories from an existing interface, to determine the position of the next interface. The methods differ  in the way in which the information from these trial trajectories is used. In the trial interface method, a ``trial'' interface is placed, the probability of reaching this interface is estimated, and the trial interface is shifted until the estimated probability lies within an acceptable range. This method is very simple to implement in an existing DFFS code, because the information needed from the trial runs (simply whether they succeeded or failed) is the same as in a conventional FFS simulation. The interface shifting step can easily be parallelized and information from some of the trial runs can be re-used in the actual FFS step once the interface has been fixed. The exploring scouts method is in some ways more sophisticated: here, trial runs are fired from the existing interface and the distribution of 
the maximum $\lambda$ values which 
they reach is used to determine a position for the next interface which corresponds to the target probability. This method has the advantage that one knows {\em{a priori}} how many trial runs will be needed to fix the interface position (important in some parallel implementations of FFS) and that the maximum length of these trial runs is fixed (albeit with some loss of accuracy in the interface position if the runs are too short).  This may be important for problems where trial runs require many computational steps (e.g. free energy barriers that are not sharply peaked, or where returning to the initial state happens slowly). However, implementation of the exploring scouts method requires slightly more modifications to existing DFFS codes, since one needs to know the maximal value of $\lambda$ reached by the trial runs, rather than their success or failure, as in standard DFFS.
While we did not test it here, one could of course combine the trial interface and exploring scouts methods within a single DFFS run, for example estimating the position of a trial interface using exploring scouts, then, in a second step, firing trial runs to the trial interface to check whether its probability is acceptable. 

We have demonstrated the use of both methods, for a simple example of a single particle in a one-dimensional potential (where we showed that the computational efficiency indeed agrees well with our theoretical predictions), and for the more realistic example of the crystallization of Yukawa particles, a computationally intensive system where the nucleation free energy barrier is {\em{a priori}} unknown. In the latter case, automatic interface placement led to a large saving in both user and computational time, compared to manual interface placement, even when the manual placement uses some prior knowledge of the shape of the free energy landscape.

The methods presented here should greatly improve the feasibility and computational efficiency of DFFS simulations for computationally expensive systems. Of course, our methods and the approach of Borrero and Escobedo \cite{borrero} are not incompatible: having placed a set of interfaces automatically using either the trial interface method or the exploring scouts method, one can further optimize their placement iteratively via the method of Borrero and Escobedo, if necessary. For the rare event problem tested here (the crystallization of Yukawa particles), we found that this did not result in any further improvement.

Our focus here has been on automatic interface placement for direct FFS (DFFS) simulations, in which the entire ensemble of trajectories is propagated forward in order parameter space, one interface at a time. In other variants of the FFS method (e.g. branched growth, Rosenbluth-like sampling~\cite{FFS2,ffs,escobedorev}, S-PRES \cite{spres} or NS-FFS \cite{nsffs}), transition paths from initial to final state are instead  generated in a one-at-a-time fashion. It should be possible to develop modifications of the automatic interface placement methods for these FFS variants: for example applying either the trial interface or exploring scouts method to fix the interfaces during the generation of the first transition path. The approaches presented here should also be compatible with other interface-based rare event simulation methods such as transition interface sampling~\cite{vanerp1,vanerp2}. Finally we note that the methods described here could be extended to interfaces that depend on more than one 
order parameter.  For example, in the exploring scouts method, one might track the trajectories of the scouts in two coordinates and set the interfaces to be optimal lines in the space of these coordinates. As well as optimising interface placement, this could also provide a way to adjust the choice of reaction coordinate, on-the-fly during an FFS simulation.

The methods described in this paper have already been implemented in the parallel rare event simulation framework FRESHS~\cite{freshs}, which allows the generic use of both FFS and other rare event simulation methods. This framework will soon be publicly available as an open-source package.

\section*{Acknowledgments}
The authors thank Kevin Stratford, Juho Lintuvuori and Chantal Valeriani for helpful discussions. 
R.J.A. was funded by a Royal Society University Research Fellowship and by EPSRC under grant EPSRC/EP/I030298/1. A.A. and K.K. were funded by the cluster of excellence ``SimTech'', University of Stuttgart.

\begin{appendix}

\section{Optimal flux calculations}\label{sec:ofc}

Here we describe in more detail our theoretical analysis of the computational efficiency of DFFS, and present our analytical expression for the efficiency as a function of the transition probability. We assume that the transition probability $p_i=p$ is the same for all interfaces. In contrast to the work of Borrero and Escobedo \cite{borrero}, we do not fix the number $n$ of interfaces. Instead, $n$ is determined by the relation 
\begin{equation}
 n = \frac{\log{P_B}}{\log{p}}.
\end{equation}
where $P_B = \prod p_i$ is the probability that a trajectory leaving $A$ reaches $B$ before returning to $A$. Following \cite{efficiency}, we define the computational efficiency as
\begin{equation}\label{eq:eff2}
\mathcal{E} = \frac{1}{\mathcal{C}\mathcal{V}}
\end{equation}
where $\mathcal {C}$  and  $\mathcal {V}$ represent the computational cost of an FFS calculation, and the statistical error (variance) in the resulting rate constant measurement. We use the expressions for  $\mathcal {C}$  and  $\mathcal {V}$ derived in \cite{efficiency} to predict the dependence of $\mathcal{E}$ on the transition probability $p$. 

\subsection{Computational cost}
The computational cost of a DFFS calculation, in simulation steps, is approximated in \cite{efficiency} by 
\begin{equation}
  \mathcal{C} \approx N_0 R + M\sum_{i=1}^{n-1}{C_i}
 \label{eq:cost}
\end{equation}
where $N_0$ is the number of configurations stored at $\lambda_A$, $R$ is the cost of generating each of these configurations, $M$ is the number of trials per interface (note we have assumed this to be constant) and $C_i$ is the average cost of firing a trial run from interface $i$. Note that Eq.(\ref{eq:cost}) describes the total cost of the FFS run rather than the cost per configuration at $\lambda_A$, as in ref.~\cite{efficiency}. Simplifying somewhat the calculation in \cite{efficiency}, we assume that the cost of a trial run is linearly proportional to the number of interfaces that it crosses, with proportionality constant $S/n$ (since the spacing between interfaces is inversely proportional to $n$; thus $S$ is the cost of a trajectory from $A$ to $B$). Thus a trial  run from $\lambda_i$ to $\lambda_{i+1}$ has cost $S/n$ while a run from $\lambda_i$ to $\lambda_A$ has cost $iS/n$. Taking into account the relative probabilities of these outcomes gives 
\begin{equation}
 C_i \approx \frac{S}{n}[p + i(1-p)]
 \label{eq:ci}
\end{equation}
resulting in the following expression for the cost:
\begin{equation}
\begin{split}
  \mathcal{C} \approx  N_0R+\frac{SM}{n} \sum_{i=1}^{n-1}{[p + i(1-p)]} \\ = N_0\left(R+\frac{Sk}{2n} \left[2p(n-1) + n(n-1)(1-p) \right]\right)
\end{split}
 \label{eq:costnew}
\end{equation}
where $k \equiv M/N_0$. The first result in Eq.(\ref{eq:costnew}) is identical to Eq.(\ref{eq:cost}) in the main text. Substituting in the expression for $n$ in terms of $P_B$ we obtain an expression for the cost in terms of $p$ and $P_B$:
\begin{equation}
 \begin{split}
   \mathcal{C}=\frac{N_0}{2\log{p}\log{P_B}}\cdot [2R\log{P_B}\log{p} \\
   + Sk ( 3p\log{P_B}\log{p} + \log{P_B}^2 \\
   - p\log{P_B}^2-2\log{p}^2-\log{P_B}\log{p})].
\end{split}
 \label{eq:costpb}
\end{equation}

\subsection{Statistical error}
The statistical error -- i.e. the variance in a calculation of the rate constant by DFFS, is approximated as in \cite{efficiency}, by
\begin{equation}
 \mathcal{V}\approx \sum_{i=1}^{n-1}\frac{(1-p_i)}{p_iM_i}
\end{equation}
Setting $p_i=p$ and $M_i=M$ we obtain
\begin{equation}
\mathcal{ V}=\frac{1}{Mp}(n-1)(1-p)=\frac{(1-p)}{N_0kp}\left( \frac{\log{P_B}}{\log{p}} - 1 \right).
\label{eq:varpb}
\end{equation}

\subsection{Efficiency}
Bringing together Eqs.\eqref{eq:eff2}, \eqref{eq:costpb} and \eqref{eq:varpb}, we obtain the following expression for the computational efficiency in terms of $p$ and $P_B$:
\begin{equation}
 \begin{split}
  \mathcal{E} =  ( 2kp\log{P_B}\log{p}^2) \cdot [ (p-1)(\log{P_B}-\log{p}) \\
  \cdot (\log{P_B}\log{p}(Sk(1-3p)-2R) \\ + Sk
  \log{P_B}^2(p-1)+2Skp\log{p}^2)]^{-1}
 \end{split}
 \label{eq:eff}
\end{equation}
Expression (\ref{eq:eff}) was used to generate the data shown in Figure \ref{fig:cvem}.

\end{appendix}

\newpage


\begin{thebibliography}{42}
\expandafter\ifx\csname natexlab\endcsname\relax\def\natexlab#1{#1}\fi
\expandafter\ifx\csname bibnamefont\endcsname\relax
  \def\bibnamefont#1{#1}\fi
\expandafter\ifx\csname bibfnamefont\endcsname\relax
  \def\bibfnamefont#1{#1}\fi
\expandafter\ifx\csname citenamefont\endcsname\relax
  \def\citenamefont#1{#1}\fi
\expandafter\ifx\csname url\endcsname\relax
  \def\url#1{\texttt{#1}}\fi
\expandafter\ifx\csname urlprefix\endcsname\relax\def\urlprefix{URL }\fi
\providecommand{\bibinfo}[2]{#2}
\providecommand{\eprint}[2][]{\url{#2}}

\bibitem[{\citenamefont{Torrie and Valleau}(1974)}]{umbrella1}
\bibinfo{author}{\bibfnamefont{G.~M.} \bibnamefont{Torrie}} \bibnamefont{and}
  \bibinfo{author}{\bibfnamefont{J.~P.} \bibnamefont{Valleau}},
  \bibinfo{journal}{Chem. Phys. Lett.} \textbf{\bibinfo{volume}{28}},
  \bibinfo{pages}{578} (\bibinfo{year}{1974}).

\bibitem[{\citenamefont{Frenkel and Smit}(2002)}]{daan}
\bibinfo{author}{\bibfnamefont{D.}~\bibnamefont{Frenkel}} \bibnamefont{and}
  \bibinfo{author}{\bibfnamefont{B.}~\bibnamefont{Smit}},
  \emph{\bibinfo{title}{Understanding Molecular Simulation. From Algorithms to
  Applications}} (\bibinfo{publisher}{Academic Press},
  \bibinfo{address}{Boston}, \bibinfo{year}{2002}), \bibinfo{edition}{2nd} ed.

\bibitem[{\citenamefont{Chandler}(1978)}]{bencha1}
\bibinfo{author}{\bibfnamefont{D.}~\bibnamefont{Chandler}},
  \bibinfo{journal}{J. Chem. Phys.} \textbf{\bibinfo{volume}{68}},
  \bibinfo{pages}{2959} (\bibinfo{year}{1978}).

\bibitem[{\citenamefont{Bennett}(1977)}]{bencha2}
\bibinfo{author}{\bibfnamefont{C.~H.} \bibnamefont{Bennett}}, in
  \emph{\bibinfo{booktitle}{Algorithms for Chemical Computations, ACS
  Symposium, Series No.46}}, edited by
  \bibinfo{editor}{\bibnamefont{R.Christofferson}}
  (\bibinfo{publisher}{American Chemical Society},
  \bibinfo{address}{Washington, D.C.}, \bibinfo{year}{1977}).

\bibitem[{\citenamefont{Dellago et~al.}(1998)\citenamefont{Dellago, Bolhuis,
  Csajka, and Chandler}}]{dellago1}
\bibinfo{author}{\bibfnamefont{C.}~\bibnamefont{Dellago}},
  \bibinfo{author}{\bibfnamefont{P.~G.} \bibnamefont{Bolhuis}},
  \bibinfo{author}{\bibfnamefont{F.~S.} \bibnamefont{Csajka}},
  \bibnamefont{and} \bibinfo{author}{\bibfnamefont{D.}~\bibnamefont{Chandler}},
  \bibinfo{journal}{J. Chem. Phys.} \textbf{\bibinfo{volume}{108}},
  \bibinfo{pages}{1964} (\bibinfo{year}{1998}).

\bibitem[{\citenamefont{Dellago et~al.}(2002)\citenamefont{Dellago, Bolhuis,
  and Geissler}}]{dellago2}
\bibinfo{author}{\bibfnamefont{C.}~\bibnamefont{Dellago}},
  \bibinfo{author}{\bibfnamefont{P.~G.} \bibnamefont{Bolhuis}},
  \bibnamefont{and} \bibinfo{author}{\bibfnamefont{P.~L.}
  \bibnamefont{Geissler}}, \bibinfo{journal}{Adv. Chem. Phys.}
  \textbf{\bibinfo{volume}{123}}, \bibinfo{pages}{1} (\bibinfo{year}{2002}).

\bibitem[{\citenamefont{Bolhuis et~al.}(2002)\citenamefont{Bolhuis, Chandler,
  Dellago, and Geissler}}]{bolhuis_arpc}
\bibinfo{author}{\bibfnamefont{P.~G.} \bibnamefont{Bolhuis}},
  \bibinfo{author}{\bibfnamefont{D.}~\bibnamefont{Chandler}},
  \bibinfo{author}{\bibfnamefont{C.}~\bibnamefont{Dellago}}, \bibnamefont{and}
  \bibinfo{author}{\bibfnamefont{P.~L.} \bibnamefont{Geissler}},
  \bibinfo{journal}{Annu. Rev. Phys. Chem.} \textbf{\bibinfo{volume}{53}},
  \bibinfo{pages}{291} (\bibinfo{year}{2002}).

\bibitem[{\citenamefont{van Erp et~al.}(2003)\citenamefont{van Erp, Moroni, and
  Bolhuis}}]{vanerp1}
\bibinfo{author}{\bibfnamefont{T.~S.} \bibnamefont{van Erp}},
  \bibinfo{author}{\bibfnamefont{D.}~\bibnamefont{Moroni}}, \bibnamefont{and}
  \bibinfo{author}{\bibfnamefont{P.~G.} \bibnamefont{Bolhuis}},
  \bibinfo{journal}{J. Chem. Phys.} \textbf{\bibinfo{volume}{118}},
  \bibinfo{pages}{7762} (\bibinfo{year}{2003}).

\bibitem[{\citenamefont{van Erp and Bolhuis}(2005)}]{vanerp2}
\bibinfo{author}{\bibfnamefont{T.~S.} \bibnamefont{van Erp}} \bibnamefont{and}
  \bibinfo{author}{\bibfnamefont{P.~G.} \bibnamefont{Bolhuis}},
  \bibinfo{journal}{J. Comp. Phys.} \textbf{\bibinfo{volume}{205}},
  \bibinfo{pages}{157} (\bibinfo{year}{2005}).

\bibitem[{\citenamefont{van Erp}(2008)}]{vanerp2008}
\bibinfo{author}{\bibfnamefont{T.~S.} \bibnamefont{van Erp}},
  \bibinfo{journal}{Comp. Phys. Commun.} \textbf{\bibinfo{volume}{179}},
  \bibinfo{pages}{34} (\bibinfo{year}{2008}).

\bibitem[{\citenamefont{Faradjian and Elber}(2004)}]{faradjian04}
\bibinfo{author}{\bibfnamefont{A.~K.} \bibnamefont{Faradjian}}
  \bibnamefont{and} \bibinfo{author}{\bibfnamefont{R.}~\bibnamefont{Elber}},
  \bibinfo{journal}{J. Chem. Phys.} \textbf{\bibinfo{volume}{120}},
  \bibinfo{pages}{10880} (\bibinfo{year}{2004}).

\bibitem[{\citenamefont{West et~al.}(2007)\citenamefont{West, Elber, and
  Shalloway}}]{west07}
\bibinfo{author}{\bibfnamefont{A.~M.~A.} \bibnamefont{West}},
  \bibinfo{author}{\bibfnamefont{R.}~\bibnamefont{Elber}}, \bibnamefont{and}
  \bibinfo{author}{\bibfnamefont{D.}~\bibnamefont{Shalloway}},
  \bibinfo{journal}{J. Chem. Phys.} \textbf{\bibinfo{volume}{126}},
  \bibinfo{pages}{145104} (\bibinfo{year}{2007}).

\bibitem[{\citenamefont{Vanden-Eijnden
  et~al.}(2008)\citenamefont{Vanden-Eijnden, Venturoli, Ciccotti, and
  Elber}}]{ve08}
\bibinfo{author}{\bibfnamefont{E.}~\bibnamefont{Vanden-Eijnden}},
  \bibinfo{author}{\bibfnamefont{M.}~\bibnamefont{Venturoli}},
  \bibinfo{author}{\bibfnamefont{G.}~\bibnamefont{Ciccotti}}, \bibnamefont{and}
  \bibinfo{author}{\bibfnamefont{R.}~\bibnamefont{Elber}}, \bibinfo{journal}{J.
  Chem. Phys.} \textbf{\bibinfo{volume}{129}}, \bibinfo{pages}{174102}
  (\bibinfo{year}{2008}).

\bibitem[{\citenamefont{Henkelman et~al.}(2000)\citenamefont{Henkelman,
  Uberuaga, and Jonsson}}]{henkelman1}
\bibinfo{author}{\bibfnamefont{G.}~\bibnamefont{Henkelman}},
  \bibinfo{author}{\bibfnamefont{B.~P.} \bibnamefont{Uberuaga}},
  \bibnamefont{and} \bibinfo{author}{\bibfnamefont{H.}~\bibnamefont{Jonsson}},
  \bibinfo{journal}{J. Chem. Phys.} \textbf{\bibinfo{volume}{113}},
  \bibinfo{pages}{9901} (\bibinfo{year}{2000}).

\bibitem[{\citenamefont{Henkelman and Jonsson}(2000)}]{henkelman2}
\bibinfo{author}{\bibfnamefont{G.}~\bibnamefont{Henkelman}} \bibnamefont{and}
  \bibinfo{author}{\bibfnamefont{H.}~\bibnamefont{Jonsson}},
  \bibinfo{journal}{J. Chem. Phys.} \textbf{\bibinfo{volume}{113}},
  \bibinfo{pages}{9978} (\bibinfo{year}{2000}).

\bibitem[{\citenamefont{E et~al.}(2002)\citenamefont{E, Ren, and
  Vanden-Eijnden}}]{Ren02}
\bibinfo{author}{\bibfnamefont{W.}~\bibnamefont{E}},
  \bibinfo{author}{\bibfnamefont{W.}~\bibnamefont{Ren}}, \bibnamefont{and}
  \bibinfo{author}{\bibfnamefont{E.}~\bibnamefont{Vanden-Eijnden}},
  \bibinfo{journal}{Phys. Rev. B} \textbf{\bibinfo{volume}{66}},
  \bibinfo{pages}{052301} (\bibinfo{year}{2002}).

\bibitem[{\citenamefont{E et~al.}(2005)\citenamefont{E, Ren, and
  Vanden-Eijnden}}]{e_05}
\bibinfo{author}{\bibfnamefont{W.}~\bibnamefont{E}},
  \bibinfo{author}{\bibfnamefont{W.}~\bibnamefont{Ren}}, \bibnamefont{and}
  \bibinfo{author}{\bibfnamefont{E.}~\bibnamefont{Vanden-Eijnden}},
  \bibinfo{journal}{J. Phys. Chem. B} \textbf{\bibinfo{volume}{109}},
  \bibinfo{pages}{6688} (\bibinfo{year}{2005}).

\bibitem[{\citenamefont{Huber and Kim}(1996)}]{huber1996}
\bibinfo{author}{\bibfnamefont{G.~A.} \bibnamefont{Huber}} \bibnamefont{and}
  \bibinfo{author}{\bibfnamefont{S.}~\bibnamefont{Kim}},
  \bibinfo{journal}{Biophys. J.} \textbf{\bibinfo{volume}{70}},
  \bibinfo{pages}{97} (\bibinfo{year}{1996}).

\bibitem[{\citenamefont{Allen et~al.}(2005)\citenamefont{Allen, Warren, and ten
  Wolde}}]{fFFS}
\bibinfo{author}{\bibfnamefont{R.~J.} \bibnamefont{Allen}},
  \bibinfo{author}{\bibfnamefont{P.~B.} \bibnamefont{Warren}},
  \bibnamefont{and} \bibinfo{author}{\bibfnamefont{P.~R.} \bibnamefont{ten
  Wolde}}, \bibinfo{journal}{Phys. Rev. Lett.} \textbf{\bibinfo{volume}{94}},
  \bibinfo{pages}{018104} (\bibinfo{year}{2005}).

\bibitem[{\citenamefont{Allen et~al.}(2006{\natexlab{a}})\citenamefont{Allen,
  Frenkel, and ten Wolde}}]{FFS2}
\bibinfo{author}{\bibfnamefont{R.~J.} \bibnamefont{Allen}},
  \bibinfo{author}{\bibfnamefont{D.}~\bibnamefont{Frenkel}}, \bibnamefont{and}
  \bibinfo{author}{\bibfnamefont{P.~R.} \bibnamefont{ten Wolde}},
  \bibinfo{journal}{J. Chem. Phys.} \textbf{\bibinfo{volume}{124}},
  \bibinfo{pages}{024102} (\bibinfo{year}{2006}{\natexlab{a}}).

\bibitem[{\citenamefont{Allen et~al.}(2006{\natexlab{b}})\citenamefont{Allen,
  Frenkel, and ten Wolde}}]{efficiency}
\bibinfo{author}{\bibfnamefont{R.~J.} \bibnamefont{Allen}},
  \bibinfo{author}{\bibfnamefont{D.}~\bibnamefont{Frenkel}}, \bibnamefont{and}
  \bibinfo{author}{\bibfnamefont{P.~R.} \bibnamefont{ten Wolde}},
  \bibinfo{journal}{J. Chem. Phys.} \textbf{\bibinfo{volume}{124}},
  \bibinfo{eid}{194111} (\bibinfo{year}{2006}{\natexlab{b}}).

\bibitem[{\citenamefont{Allen et~al.}(2009)\citenamefont{Allen, Valeriani, and
  ten Wolde}}]{ffs}
\bibinfo{author}{\bibfnamefont{R.~J.} \bibnamefont{Allen}},
  \bibinfo{author}{\bibfnamefont{C.}~\bibnamefont{Valeriani}},
  \bibnamefont{and} \bibinfo{author}{\bibfnamefont{P.~R.} \bibnamefont{ten
  Wolde}}, \bibinfo{journal}{Journal of Physics: Condensed Matter}
  \textbf{\bibinfo{volume}{21}}, \bibinfo{pages}{463102}
  (\bibinfo{year}{2009}).

\bibitem[{\citenamefont{Valeriani et~al.}(2007)\citenamefont{Valeriani, Allen,
  Morelli, Frenkel, and ten Wolde}}]{barrier}
\bibinfo{author}{\bibfnamefont{C.}~\bibnamefont{Valeriani}},
  \bibinfo{author}{\bibfnamefont{R.~J.} \bibnamefont{Allen}},
  \bibinfo{author}{\bibfnamefont{M.~J.} \bibnamefont{Morelli}},
  \bibinfo{author}{\bibfnamefont{D.}~\bibnamefont{Frenkel}}, \bibnamefont{and}
  \bibinfo{author}{\bibfnamefont{P.~R.} \bibnamefont{ten Wolde}},
  \bibinfo{journal}{J. Chem. Phys.} \textbf{\bibinfo{volume}{127}},
  \bibinfo{pages}{114109} (\bibinfo{year}{2007}).

\bibitem[{\citenamefont{Borrero and Escobedo}(2009)}]{borrero2009}
\bibinfo{author}{\bibfnamefont{E.~E.} \bibnamefont{Borrero}} \bibnamefont{and}
  \bibinfo{author}{\bibfnamefont{F.~A.} \bibnamefont{Escobedo}},
  \bibinfo{journal}{J. Phys. Chem. B} \textbf{\bibinfo{volume}{113}},
  \bibinfo{pages}{6434} (\bibinfo{year}{2009}).

\bibitem[{\citenamefont{Borrero and Escobedo}(2007)}]{borrero2007}
\bibinfo{author}{\bibfnamefont{E.~E.} \bibnamefont{Borrero}} \bibnamefont{and}
  \bibinfo{author}{\bibfnamefont{F.~A.} \bibnamefont{Escobedo}},
  \bibinfo{journal}{J. Chem. Phys.} \textbf{\bibinfo{volume}{127}},
  \bibinfo{pages}{164101} (\bibinfo{year}{2007}).

\bibitem[{\citenamefont{Dickson and Dinner}(2010)}]{dinner1}
\bibinfo{author}{\bibfnamefont{A.}~\bibnamefont{Dickson}} \bibnamefont{and}
  \bibinfo{author}{\bibfnamefont{A.~R.} \bibnamefont{Dinner}},
  \bibinfo{journal}{Annu. Rev. Phys. Chem.} \textbf{\bibinfo{volume}{61}},
  \bibinfo{pages}{441} (\bibinfo{year}{2010}).

\bibitem[{\citenamefont{Warmflash et~al.}(2007)\citenamefont{Warmflash,
  Bhimalapuram, and Dinner}}]{dinner2}
\bibinfo{author}{\bibfnamefont{A.}~\bibnamefont{Warmflash}},
  \bibinfo{author}{\bibfnamefont{P.}~\bibnamefont{Bhimalapuram}},
  \bibnamefont{and} \bibinfo{author}{\bibfnamefont{A.}~\bibnamefont{Dinner}},
  \bibinfo{journal}{J. Chem. Phys.} \textbf{\bibinfo{volume}{127}},
  \bibinfo{pages}{154112} (\bibinfo{year}{2007}).

\bibitem[{\citenamefont{Escobedo et~al.}(2009)\citenamefont{Escobedo, Borrero,
  and Araque}}]{escobedorev}
\bibinfo{author}{\bibfnamefont{F.~A.} \bibnamefont{Escobedo}},
  \bibinfo{author}{\bibfnamefont{E.~E.} \bibnamefont{Borrero}},
  \bibnamefont{and} \bibinfo{author}{\bibfnamefont{J.~C.}
  \bibnamefont{Araque}}, \bibinfo{journal}{Journal of Physics: Condensed
  Matter} \textbf{\bibinfo{volume}{21(33)}}, \bibinfo{pages}{333101}
  (\bibinfo{year}{2009}).

\bibitem[{\citenamefont{Valeriani}(2007)}]{valeriani}
\bibinfo{author}{\bibfnamefont{C.}~\bibnamefont{Valeriani}}, Ph.D. thesis,
  \bibinfo{school}{FOM AMOLF} (\bibinfo{year}{2007}).

\bibitem[{\citenamefont{Berryman and Schilling}(2010)}]{spres}
\bibinfo{author}{\bibfnamefont{J.~T.} \bibnamefont{Berryman}} \bibnamefont{and}
  \bibinfo{author}{\bibfnamefont{T.}~\bibnamefont{Schilling}},
  \bibinfo{journal}{J. Chem. Phys.} \textbf{\bibinfo{volume}{133}},
  \bibinfo{pages}{244101} (\bibinfo{year}{2010}).

\bibitem[{\citenamefont{Becker et~al.}(2012)\citenamefont{Becker, Allen, and
  ten Wolde}}]{nsffs}
\bibinfo{author}{\bibfnamefont{N.~B.} \bibnamefont{Becker}},
  \bibinfo{author}{\bibfnamefont{R.~J.} \bibnamefont{Allen}}, \bibnamefont{and}
  \bibinfo{author}{\bibfnamefont{P.~R.} \bibnamefont{ten Wolde}},
  \bibinfo{journal}{J. Chem. Phys.} \textbf{\bibinfo{volume}{136}},
  \bibinfo{eid}{174118} (\bibinfo{year}{2012}).


\bibitem[{\citenamefont{Sear}(2008)\citenamefont{Sear}}]{sear}
\bibinfo{author}{\bibfnamefont{R.~P.} \bibnamefont{Sear}},
  \bibinfo{journal}{J. Chem. Phys.} \textbf{\bibinfo{volume}{128}},
  \bibinfo{eid}{214513} (\bibinfo{year}{2008}).

\bibitem[{\citenamefont{Van Erp}(2012)\citenamefont{Van Erp}}]{vanerp}
\bibinfo{author}{\bibfnamefont{T.~S.} \bibnamefont{van Erp}},
  \bibinfo{journal}{Adv. Chem. Phys.} \textbf{\bibinfo{volume}{151}},
  \bibinfo{pages}{27} (\bibinfo{year}{2012}).



\bibitem[{\citenamefont{Filion et~al.}(2010)\citenamefont{Filion et al}}]{dijkstra}
\bibinfo{author}{\bibfnamefont{L.} \bibnamefont{Filion}},
  \bibinfo{author}{\bibfnamefont{M.} \bibnamefont{Hermes}}
   \bibinfo{author}{\bibfnamefont{R.} \bibnamefont{Ni}}, \bibnamefont{and}
  \bibinfo{author}{\bibfnamefont{M.} \bibnamefont{Dijkstra}},
  \bibinfo{journal}{J. Chem. Phys.} \textbf{\bibinfo{volume}{133}},
  \bibinfo{eid}{244115} (\bibinfo{year}{2010}).

\bibitem[{\citenamefont{Valeriani et~al.}(2005)\citenamefont{Valeriani et al}}]{valeriani_nuc}
\bibinfo{author}{\bibfnamefont{C.} \bibnamefont{Valeriani}},
  \bibinfo{author}{\bibfnamefont{E.} \bibnamefont{Sanz}}, \bibnamefont{and}
  \bibinfo{author}{\bibfnamefont{D.} \bibnamefont{Frenkel}},
  \bibinfo{journal}{J. Chem. Phys.} \textbf{\bibinfo{volume}{122}},
  \bibinfo{eid}{194501} (\bibinfo{year}{2005}).


\bibitem[{\citenamefont{Velez-Vega et~al.}(2010)\citenamefont{Velez-Vega et al}}]{vega}
\bibinfo{author}{\bibfnamefont{C.} \bibnamefont{Velez-Vega}},
  \bibinfo{author}{\bibfnamefont{E.~E.} \bibnamefont{Borrero}}, \bibnamefont{and}
  \bibinfo{author}{\bibfnamefont{F.~A.} \bibnamefont{Escobedo}},
  \bibinfo{journal}{J. Chem. Phys.} \textbf{\bibinfo{volume}{133}},
  \bibinfo{eid}{105103} (\bibinfo{year}{2010}).

\bibitem[{\citenamefont{Borrero and Escobedo}(2006)\citenamefont{Borrero and Escobedo}}]{borrero_lattice}
  \bibinfo{author}{\bibfnamefont{E.~E.} \bibnamefont{Borrero}}, \bibnamefont{and}
  \bibinfo{author}{\bibfnamefont{F.~A.} \bibnamefont{Escobedo}},
  \bibinfo{journal}{J. Chem. Phys.} \textbf{\bibinfo{volume}{125}},
  \bibinfo{eid}{164904} (\bibinfo{year}{2006}).

\bibitem[{\citenamefont{Borrero and Escobedo}(2008)}]{borrero}
\bibinfo{author}{\bibfnamefont{E.~E.} \bibnamefont{Borrero}} \bibnamefont{and}
  \bibinfo{author}{\bibfnamefont{F.~A.} \bibnamefont{Escobedo}},
  \bibinfo{journal}{J. Chem. Phys.} \textbf{\bibinfo{volume}{129}},
  \bibinfo{eid}{024115} (\bibinfo{year}{2008}).

\bibitem[{\citenamefont{Borrero et~al.}(2011)\citenamefont{Borrero, Weinwurm,
  and Dellago}}]{borrero2}
\bibinfo{author}{\bibfnamefont{E.}~\bibnamefont{Borrero}},
  \bibinfo{author}{\bibfnamefont{M.}~\bibnamefont{Weinwurm}}, \bibnamefont{and}
  \bibinfo{author}{\bibfnamefont{C.}~\bibnamefont{Dellago}},
  \bibinfo{journal}{The Journal of chemical physics}
  \textbf{\bibinfo{volume}{134}}, \bibinfo{pages}{244118}
  (\bibinfo{year}{2011}).

\bibitem[{\citenamefont{C\'erou and Guyader}(2007)}]{ams}
\bibinfo{author}{\bibfnamefont{F.}~\bibnamefont{C\'erou}} \bibnamefont{and}
  \bibinfo{author}{\bibfnamefont{A.}~\bibnamefont{Guyader}},
  \bibinfo{journal}{Stochastic Analysis and Applications}
  \textbf{\bibinfo{volume}{25}}, \bibinfo{pages}{417} (\bibinfo{year}{2007}).

\bibitem[{\citenamefont{Weeks et~al.}(1971)\citenamefont{Weeks, Chandler, and
  Andersen}}]{weeks71a}
\bibinfo{author}{\bibfnamefont{J.~D.} \bibnamefont{Weeks}},
  \bibinfo{author}{\bibfnamefont{D.}~\bibnamefont{Chandler}}, \bibnamefont{and}
  \bibinfo{author}{\bibfnamefont{H.~C.} \bibnamefont{Andersen}},
  \bibinfo{journal}{J. Chem. Phys.} \textbf{\bibinfo{volume}{54}},
  \bibinfo{pages}{5237} (\bibinfo{year}{1971}).

\bibitem[{\citenamefont{Auer and Frenkel}(2002)}]{yukawa}
\bibinfo{author}{\bibfnamefont{S.}~\bibnamefont{Auer}} \bibnamefont{and}
  \bibinfo{author}{\bibfnamefont{D.}~\bibnamefont{Frenkel}},
  \bibinfo{journal}{Journal of Physics: Condensed Matter}
  \textbf{\bibinfo{volume}{14}}, \bibinfo{pages}{7667} (\bibinfo{year}{2002}).

\bibitem[{\citenamefont{Sanz et~al.}(2007)\citenamefont{Sanz, Valeriani,
  Frenkel, and Dijkstra}}]{sanz2007}
\bibinfo{author}{\bibfnamefont{E.}~\bibnamefont{Sanz}},
  \bibinfo{author}{\bibfnamefont{C.}~\bibnamefont{Valeriani}},
  \bibinfo{author}{\bibfnamefont{D.}~\bibnamefont{Frenkel}}, \bibnamefont{and}
  \bibinfo{author}{\bibfnamefont{M.}~\bibnamefont{Dijkstra}},
  \bibinfo{journal}{Phys. Rev. Lett.} \textbf{\bibinfo{volume}{99}},
  \bibinfo{pages}{055501} (\bibinfo{year}{2007}).

\bibitem[{\citenamefont{Azhar et~al.}(2000)\citenamefont{Azhar, Baus, Ryckaert,
  and Meijer}}]{azhar}
\bibinfo{author}{\bibfnamefont{F.~E.} \bibnamefont{Azhar}},
  \bibinfo{author}{\bibfnamefont{M.}~\bibnamefont{Baus}},
  \bibinfo{author}{\bibfnamefont{J.-P.} \bibnamefont{Ryckaert}},
  \bibnamefont{and} \bibinfo{author}{\bibfnamefont{E.~J.}
  \bibnamefont{Meijer}}, \bibinfo{journal}{J. Chem. Phys.}
  \textbf{\bibinfo{volume}{112}}, \bibinfo{pages}{5121} (\bibinfo{year}{2000}).

\bibitem[{\citenamefont{Limbach et~al.}(2006)\citenamefont{Limbach, Arnold,
  Mann, and Holm}}]{espresso}
\bibinfo{author}{\bibfnamefont{H.-J.} \bibnamefont{Limbach}},
  \bibinfo{author}{\bibfnamefont{A.}~\bibnamefont{Arnold}},
  \bibinfo{author}{\bibfnamefont{B.~A.} \bibnamefont{Mann}}, \bibnamefont{and}
  \bibinfo{author}{\bibfnamefont{C.}~\bibnamefont{Holm}},
  \bibinfo{journal}{Comput. Phys. Commun. 174} \textbf{\bibinfo{volume}{9}},
  \bibinfo{pages}{704} (\bibinfo{year}{2006}).

\bibitem[{\citenamefont{Kratzer et~al.}(in preparation)\citenamefont{Kratzer,
  Berryman, Taudt, and Arnold}}]{freshs}
\bibinfo{author}{\bibfnamefont{K.}~\bibnamefont{Kratzer}},
  \bibinfo{author}{\bibfnamefont{J.~T.} \bibnamefont{Berryman}},
  \bibinfo{author}{\bibfnamefont{A.}~\bibnamefont{Taudt}}, \bibnamefont{and}
  \bibinfo{author}{\bibfnamefont{A.}~\bibnamefont{Arnold}} (\bibinfo{year}{in
  preparation}).

\bibitem[{\citenamefont{Lechner and Dellago}(2008)}]{q6}
\bibinfo{author}{\bibfnamefont{W.}~\bibnamefont{Lechner}} \bibnamefont{and}
  \bibinfo{author}{\bibfnamefont{C.}~\bibnamefont{Dellago}},
  \bibinfo{journal}{J. Chem. Phys.} \textbf{\bibinfo{volume}{129}},
  \bibinfo{eid}{114707} (\bibinfo{year}{2008}).

\bibitem[{\citenamefont{Kratzer and Arnold}(in preparation)}]{kratzeryuk}
\bibinfo{author}{\bibfnamefont{K.}~\bibnamefont{Kratzer}} \bibnamefont{and}
  \bibinfo{author}{\bibfnamefont{A.}~\bibnamefont{Arnold}} (\bibinfo{year}{in
  preparation}).

\end{thebibliography}

\end{document}